\documentclass[a4paper,fleqn,usenatbib]{mnras}

\usepackage{newtxtext,newtxmath}
\usepackage[T1]{fontenc}
\usepackage{ae,aecompl}

\usepackage{graphicx}	% Including figure files
\usepackage{amsmath}	% Advanced maths commands
\usepackage{amssymb}	% Extra maths symbols
\usepackage{bm}  % bold math
\usepackage{comment}
\usepackage{subfig}
 \usepackage{enumitem}
\usepackage[pdftex,dvipsnames]{xcolor}
\usepackage[linecolor=MidnightBlue,backgroundcolor=MidnightBlue!25,bordercolor=MidnightBlue,prependcaption]{todonotes}

\newcommand{\ie}{{\it i.e.}}
\newcommand{\eg}{{\it e.g.}}

\DeclareMathOperator{\asinh}{asinh}

\usepackage{ulem} % To scratch words
\definecolor{myorange}{RGB}{255, 104, 51}

\newcounter{enumCount}

\title[Sample variance in redshift calibration]{Propagating sample variance uncertainties in redshift calibration: simulations, theory and application to the COSMOS2015 data}

\author[S\'anchez, Raveri, Alarcon \& Bernstein]{
Carles S\'anchez$^{1}$\thanks{Corresponding author: carless@physics.upenn.edu},
Marco Raveri$^{1}$\thanks{Corresponding author: mraveri@sas.upenn.edu},
Alex Alarcon$^{2}$ \& %\thanks{Corresponding author: alexalarcongonzalez@gmail.com},
Gary M. Bernstein$^{1}$  
\vspace{2mm}
\\
$^{1}$ Department of Physics \& Astronomy, University of Pennsylvania, 209 S. 33rd St., Philadelphia, PA 19104, USA\\
$^{2}$ HEP Division, Argonne National Laboratory, Lemont, IL 60439, USA\\
}

\vspace{1mm}

\date{\today}

\pubyear{2020}

\begin{document}
\label{firstpage}
\pagerange{\pageref{firstpage}--\pageref{lastpage}}
\maketitle

\begin{abstract}
Cosmological analyses of galaxy surveys rely on knowledge of the redshift distribution of their galaxy sample.
This is usually derived from a spectroscopic and/or many-band photometric calibrator survey of a small patch of sky.
The uncertainties in the redshift distribution of the calibrator sample include a contribution from shot noise, or Poisson sampling errors, but, given the small volume they probe, they are dominated by sample variance introduced by large-scale structures.
Redshift uncertainties have been shown to constitute one of the leading contributions to systematic uncertainties in cosmological inferences from weak lensing and galaxy clustering, and hence they must be propagated through the analyses. 
In this work, we study the effects of sample variance on small-area redshift surveys, from theory to simulations to the 
COSMOS2015 data set.
We present a three-step Dirichlet method of resampling a given survey-based redshift 
calibration distribution  
to enable the propagation of both shot noise and sample variance uncertainties. The method can accommodate different levels of prior confidence on different redshift sources.
This method can be applied to any calibration sample with known redshifts and phenotypes (\ie~cells in a \textit{self-organizing map}, or some other way of discretizing photometric space), and provides a simple way of propagating prior redshift uncertainties into cosmological analyses.  
As a worked example, we apply the full scheme to the COSMOS2015 data set, for which we also present a new, principled SOM algorithm designed to handle noisy photometric data.  
We make available a catalog of the resulting resamplings of the COSMOS2015 galaxies.
\end{abstract}

\begin{keywords}
observational cosmology, galaxy surveys, photometric redshifts
\end{keywords}

 %%%%%%%%%%%%%%%%%%%%%%%%%%%%%%%%%%%%%%%%%%%%%%%%%%

 %%%%%%%%%%%%%%%%% BODY OF PAPER %%%%%%%%%%%%%%%%%%
% Split into different files for each section, to keep files short. We can change to everything in one file if we want to.    

\section{Introduction}
\label{sec:intro}

Imaging (or photometric) galaxy surveys, such as  
the \textit{Sloan Digital Sky Survey} \citep[\textit{SDSS},][]{York:2000gk}, 
\textit{PanSTARRS} \citep{Kaiser2000}, 
the \textit{Kilo-Degree Survey} \citep[\textit{KiDS},][]{Jong2013}, 
the \textit{Dark Energy Survey} \citep[\textit{DES},][]{Flaugher2015}, 
the \textit{Hyper-Suprime-Cam} survey \citep[\textit{HSC},][]{HSC2012}, 
or the \textit{Legacy Survey of Space and Time} \citep[\textit{LSST},][]{LSST2012},
provide key information about the large-scale structure of the Universe using the weak gravitational lensing and clustering of galaxies, and they constitute one of the most powerful probes for testing cosmological models.   

In order to perform unbiased cosmological analyses of imaging surveys it is very important to characterize the redshift distributions $n(z)=\mathrm{d}N/\mathrm{d}z\,\mathrm{d}A$ of the corresponding galaxy samples, and systematic errors in that characterization may directly lead to biases in the cosmological parameter estimation \citep{Huterer2006,Hildebrandt2012,Cunha2012,Benjamin2013,Huterer2013,Bonnett2015,Samuroff17, Hoyle2017,Hildebrandt2017}.
Relatedly, recent comparisons between cosmological parameters obtained from imaging suveys \citep{Hildebrandt2018,Troxel2018,Hikage2018} and the cosmic microwave background \citep{PlanckCollaboration2018} have claimed discrepancies of up to $3.2\sigma$ in their estimates for the amplitude of density fluctuations in a $\Lambda$CDM universe \citep{Asgari2019}. Even though such discrepancies may be attributed to a failure of the $\Lambda$CDM model \citep{Joudaki2016}, that claim would need significant evidence and thorough testing. Alternatively, some studies suggest it may instead be pointing to systematic biases in the weak lensing analysis methodologies \citep{Troxel2018a,Joudaki2019,Asgari2019,Wright2019}. Moreover, such studies indicate that a major difference in the methodologies of those analyses lies in the redshift calibration, and that this has the potential to produce such discrepancy. Redshift calibration requires substantial improvement for the success of the current and next generations of imaging surveys.      

Redshift constraints in photometric surveys usually begin with  
external data on the redshift distribution of their galaxy sample, which can be considered as a prior on $n(z)$ for subsequent survey analyses. 
Because spectroscopic or high-quality photometric redshifts are very costly in time and resources, such information typically comes from a small area on the sky and, therefore, it is subject to both shot noise and sample variance due to the large-scale structure of the Universe. 
As redshift uncertainties can dominate the error budget in current and future weak lensing analyses, it is very important to propagate such sources of uncertainty into the derived cosmological constraints. However, there is yet no clear way of sampling from that redshift prior while including shot noise and sample variance as sources of uncertainty, and hence these have been frequently overlooked or estimated relying on simplified simulated galaxy catalogs.  

There exist ways of estimating redshift distribution uncertainties from the data themselves, using subsampling methods such as bootstrapping. Such methods assume, however, that the subsamples are independent draws of a given random field, which is not true if they are correlated by large-scale structure fluctuations. 
Recently, some studies have used the Dirichlet distribution to model the information contained in the redshift calibration sample \citep{Leistedt2016,Sanchez2019,Alarcon2019}, as a way to propagate uncertainties from the prior into a redshift posterior. 
This is a good choice because the Dirichlet distribution produces samples of a distribution which preserve normalization and have positive elements, among other properties.  But again, when the redshift information comes from small patches of the sky, the Dirichlet sampling is only propagating shot noise from the calibrator survey, while if the patches are small enough, sample variance from large-scale structure may be the dominant source of uncertainty \citep{Cunha2012}. 

In this paper, we will study the problem of sample variance in redshift estimation in detail, and introduce a number of advancements on several fronts. So far, this problem has mostly been studied in simulations (\eg~\citealt{Cunha2012}). That has some drawbacks, such as the limitations in redshift range ($N$-body simulations are typically not reliable in a broad redshift range like $0<z<5$), the fixed cosmology, and the associated statistical uncertainties for the simulation volume. In this work, for the first time, we develop a theoretical estimate of the sample variance contribution to redshift uncertainties, and we validate that using $N$-body simulations. The theoretical estimation has some advantages, such as the unlimited redshift range and the possibility to explore the effects of super-sample covariance, cosmological model dependencies, or redshift-space distortions and lensing magnification. 
We then introduce a novel sampling method based on the Dirichlet distribution which takes as input the theoretical estimate of sample variance and the redshift-survey catalog, then produces samples of $n(z)$ distribution that draw from uncertainties due to both the shot noise and the sample variance in the catalogs.  This yields the correct sampling of uncertainties in the prior for $n(z)$ and hence propagation of those into cosmology analyses of a photometric survey. The method is based on the phenotype approach described in \citet{Sanchez2019,Alarcon2019}. Additionally, in our application of the method to the COSMOS2015 data sample \citep{Laigle2016}, we present a new SOM algorithm designed to handle noisy photometric data, to be used in the phenotype characterization of a galaxy population.   

This paper is organized as follows. In Section \ref{sec:sims} we describe the simulations used in this work and the phenotype approach we will work with throughout the paper. In Section \ref{sec:uncertainties} we use those simulations to characterize the shot noise and sample variance uncertainties in the redshift distribution, and write down a parameterization that separates the two contributions. We develop the sample variance contribution from a theoretical perspective in Section \ref{sec:theory}. In Section \ref{sec:method} we present a sampling scheme that can produce realizations of a redshift distribution including shot noise and sample variance uncertainties. Finally, in Section \ref{sec:results} we validate the results in simulations, and we apply them to real data in Section \ref{sec:cosmos} (using a SOM algorithm described in the Appendix). Conclusions are presented in Section \ref{sec:con}.    

\section{Framework and simulated data}
\label{sec:sims}

\begin{figure}
\centering
	\includegraphics[width=0.4\textwidth]{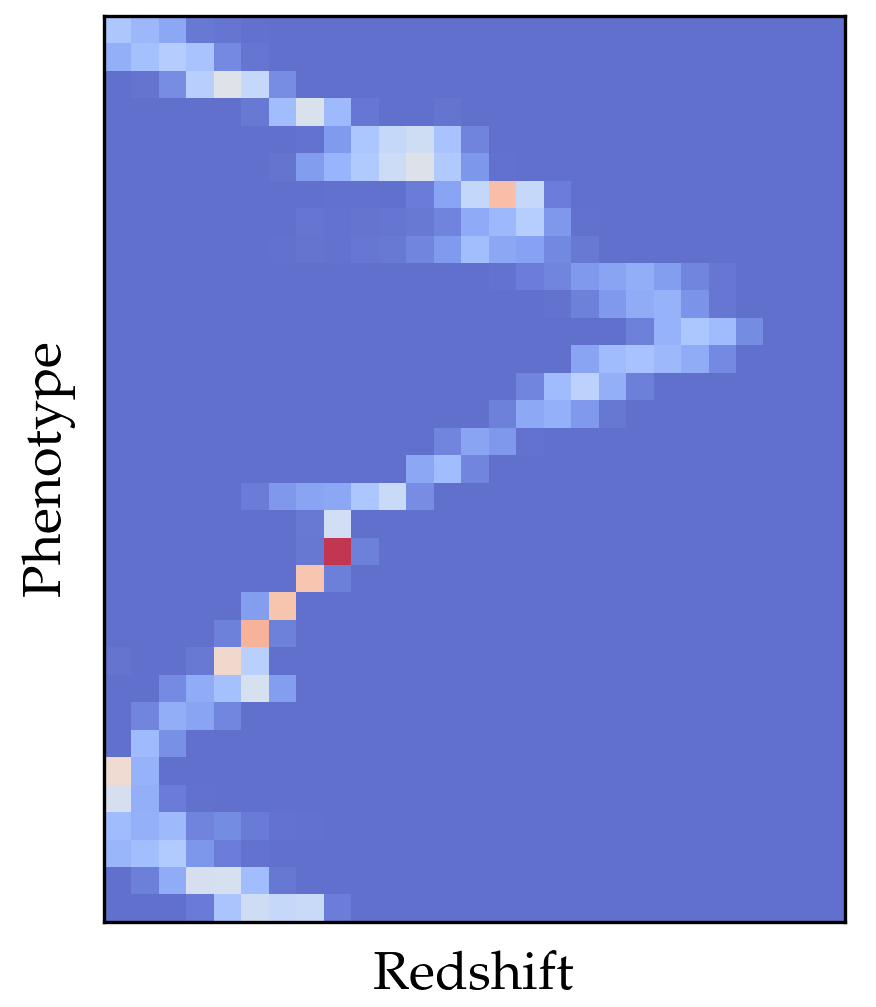}
	\caption{Graphical representation of the redshift and phenotype description of a galaxy population. The color scale is the relative density $f_{zt}$ of galaxies as each combination of redshift $z$ and phenotype $t$. Apparent is the correlation between phenotypes and redshifts: phenotypes are generally only allowed in a certain redshift range, and some phenotypes will have a tighter redshift distribution than others. This Figure shows only a subsample of phenotypes, for easier visualization. The ordering of phenotypes is arbitrary.}
\label{fig:pzt}
\end{figure}

We will work in the context of the phenotype redshift approach \citep{Sanchez2019,Alarcon2019,Buchs2019}, in which we model the galaxy population as a 2D histogram in redshift $z$ and phenotype $t$ (see Figure \ref{fig:pzt} for a graphical description), such that $p(z,t) \equiv f_{zt}$ gives the fraction of the population in each $(z,t)$ bin. $N_{zt}$ will be the counts on that histogram for a finite realized galaxy sample. For the implementation of this scheme, we will use a combination of deep survey observations and self-organizing maps (SOMs). Deep observations are often available for surveys like the Dark Energy Survey (DES), Euclid, or LSST via searches for SNe \citep{DES_DR1,Euclid_deepfield_2018,LSST_deepfield_2018}, and these provide essentially zero-noise photometric measurements and additional filters for galaxies in specific fields (henceforth deep fields, or simply DFs), and provide an empirical sampling of the distribution of galaxies in the observed photometric space. In turn, SOMs provide a data-driven way of mapping and discretizing that observed photometric space \citep{Masters2015}, so that we can use the cells in a SOM trained in the DFs as the definition of our galaxy phenotypes $t$. In addition, self-organizing maps have been extensively used for redshift studies in the past years \citep{Alarcon2019,Buchs2019,Hemmati2019,Wright2019}, and hence all the results presented in this work can be easily accomodated in all of the current redshift calibration efforts that utilize SOMs for the purpose of redshift calibration.

In this work, as in \citet{Alarcon2019}, we will use the public MICE2 simulation,\footnote{The data can be downloaded from CosmoHub \citep{COSMOHUB}, \url{https://cosmohub.pic.es/}.} a mock galaxy catalog created from a lightcone of a dark-matter-only N-body simulation that contains  $\sim$200 million galaxies over one sky octant ($\sim 5000$ deg$^{2}$) and up to $z=1.4$. The MICE2 simulation has realistic clustering properties given by a $\Lambda$CDM cosmology with parameters $\Omega_{m}=0.25$, $\Omega_b=0.044$, $h=0.7$, $n_s=0.95$, $\Omega_{\Lambda}=0.75$, $\sigma_8=0.8$ and $w=-1$. The galaxies in the simulation have realistic spectral energy distributions (SEDs) assigned from the COSMOS catalog (\citealt{COSMOSILBERT}) that reproduce the observed color-magnitude distribution as well as clustering observations as a function of colors and luminosity (see \citealt{MICE2} for more details). Once the galaxy SED is known, magnitudes are computed based on the luminosity and redshift of the galaxy. The galaxy properties, clustering and lensing in the simulation have been thoroughly validated in \cite{MICE0,MICE1,MICE2,MICE3}. 

To define phenotypes in the simulation, we create a self-organizing map on a square grid with periodic boundary conditions, similar to the SOM in \citet{Masters2015}. The SOM is trained with eight colors, defined as $\mathrm{mag}-i$, where $\mathrm{mag}=\{u,g,r,z,Y,J,H,K\}$, in a $32\times32$ grid, and presents a median redshift dispersion of 0.030 per cell. This is the same SOM as the deep SOM described in \citet{Alarcon2019}; please refer to Figure 2 in that work for a graphical representation. For the redshift part, we discretize the redshift space of the simulation in 42 redshift bins of width 0.03 in the redshift range $0.15\leq z \leq 1.41$.

\section{Shot noise and sample variance: characterizing the uncertainties}
\label{sec:uncertainties}

\begin{figure*}
\centering
	\includegraphics[width=0.9\textwidth]{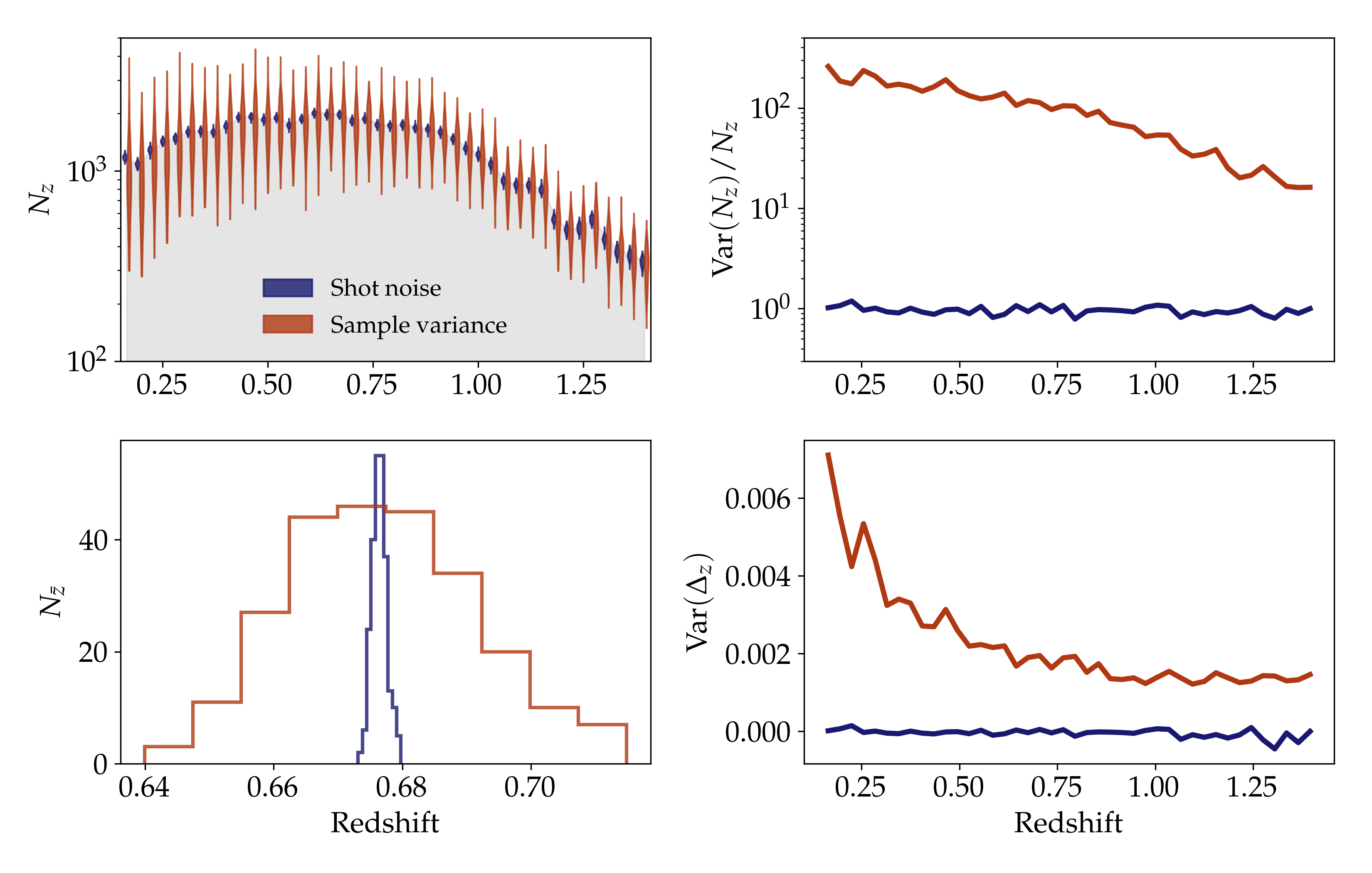}
	\caption{Effects of sample variance and shot noise in the redshift distribution of galaxies, as determined from the variation of $N_z$ among the healpixel and random-equivalent sets of simulated spectroscopic surveys, respectively.   \textit{(Upper-left panel):} Violin plots with the distribution of each redshift bin in the redshift distributions. \textit{(Lower-left panel):} Distribution of the mean redshifts of each sample redshift distribution. \textit{(Upper-right panel):} Normalized variance in the redshift distributions as a function of redshift. \textit{(Lower-right panel):} Sample variance as a function of redshift, as parametrized in Equation (\ref{param2}).  }
\label{fig:uncertainties}
\end{figure*}

The characterization of redshift distributions in large imaging surveys relies on prior information coming from smaller spectroscopic or many-band, deep photometric surveys -- we will refer to those as \textit{calibration} samples. In order to perform accurate cosmological analyses using such information, those calibration samples should be representative of the entire target population. However, that is very difficult to achieve in practice. Assuming that the calibration samples have the same selection and completeness as the entire survey, there are two statistical reasons why they will still be different than the full sample:
\begin{itemize}
	\item Shot noise: calibration samples are costly in time and resources, which makes the number of galaxies in them very small compared to the total number of galaxies in a survey. Therefore, \textit{Poisson} fluctuations are significant for them.         
	\item Sample variance: calibration samples are typically from small sky patches subject to large fluctuations due to the large-scale structure of the Universe, \ie~they may be sitting on some void or cluster of galaxies at a given $z$ and not be a fair representation of the Universe. This effect is known as sample variance, and can introduce fluctuations which are an order of magnitude larger than shot noise \citep{VanWaerbeke2006,Cunha2012}.     
\end{itemize}   

To study the sources of uncertainties that go into redshift priors we construct two sets of simulated spectroscopic redshift samples:  each sample in the first  set consist of all galaxies from one healpy sky pixel of the simulation (with nside=$2^5$), which has an area of $\sim3.5\mathrm{deg}^2$. We use 247 of those patches, with a mean number of galaxies of $\approx56000$, and standard deviation of $\approx3000.$ We will refer to these as the \textit{healpixel} samples.  Additionally, we construct a set of \textit{random-equivalent} samples \citep{Cunha2012}, by drawing 247 samples, each with 56000 galaxies drawn at random from the full simulation.  Each of the random-equivalent spectroscopic samples has the same shot noise as the previous case, but without the variance induced by large-scale structure. 
In the simulation we assume the redshift and phenotype of each galaxy are known exactly, \ie~we use the true values. In reality, redshifts will typically come from a spectroscopic or high-quality photo-$z$ sample, and phenotype will come from a SOM cell placement using deep photometry of those galaxies, so both estimates will have some additional noise. 

Next, we characterize the imprint that sample variance leaves in redshift constraints coming from calibration samples. This will allow us to develop ways to include such effects into the sampling of the redshift distribution $n(z)$, which will be the subject of Section \ref{sec:method}. We will split this Section in two parts, separating the effects of sample variance in redshift from those in phenotype.  

\subsection{Effects in redshift}

There is one critical difference between the two sources of uncertainty considered here, especially regarding their importance for redshift inference and calibration. Shot noise depends solely on the number of galaxies in a given redshift bin of the redshift sample, whereas sample variance has additional explicit dependence on redshift due to evolution of the volume elements and the large-scale clustering strength. 

We parametrize the redshift distribution of a given patch in the sky by including the contributions from shot noise (Poisson) and sample variance as follows:
\begin{equation}
	N_z = \mathrm{Poisson}[\bar N f_z (1+\Delta_z)].
	\label{param1}
\end{equation}

Here $N_z$ is the number of galaxies from the sample in redshift bin $z$, $\bar{N}$ is the angular average galaxy density, $f_z$ is the shape of the redshift distribution for the whole galaxy population and $\Delta_z$ captures variations in source density at $z$ due to the sample variance effect. If we look at the (normalized) variance of an ensemble of patches due to these effects, we find:

\begin{equation}
	\frac{\mathrm{Var}(N_z)}{\langle N_z \rangle} = 1 + N f_z \: \mathrm{Var}(\Delta_z),	
	\label{param2}
\end{equation}

where $\langle N_z \rangle$ is the average $N_z$ for different patches. The term $\mathrm{Var}(\Delta_z)$ corresponds to the contribution from sample variance alone, and is defined to not depend on the galaxy number counts.

\begin{figure*}
\centering
	\includegraphics[width=0.9\textwidth]{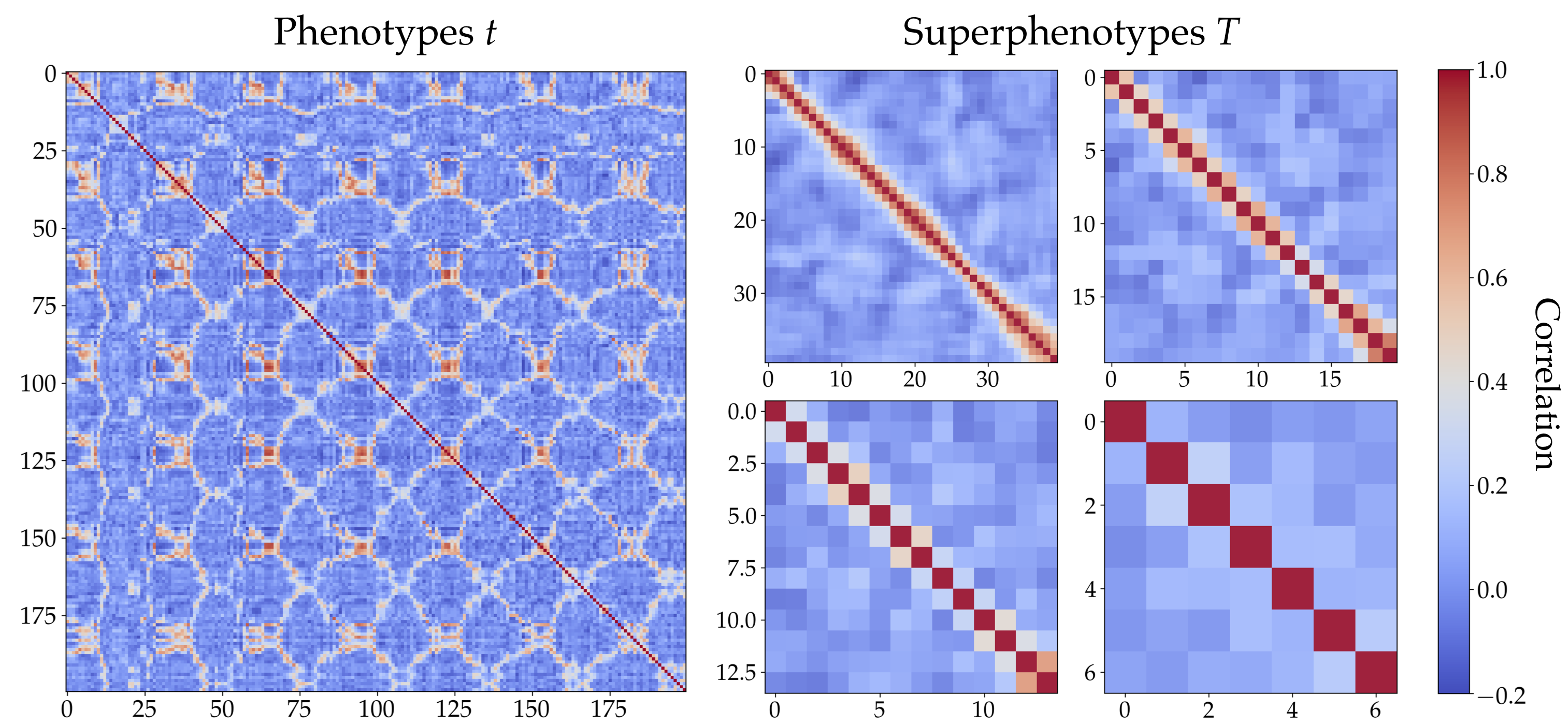}
	\caption{\textit{(Left half)}: Correlation matrix of the phenotype distribution in different patches of the simulation, showing that sample variance produces strong correlation among phenotypes. The plot shows only a subset (200 out of 1024) of phenotypes, for easier visualization. \textit{(Right half)}: Correlation matrices of superphenotypes that are constructed by joining phenotypes with mean $p(z|t)$ that lie within one, two, three and six redshift bins (left to right, top to bottom). }
\label{fig:corrs}
\end{figure*}

\begin{figure}
\centering
	\includegraphics[width=0.5\textwidth]{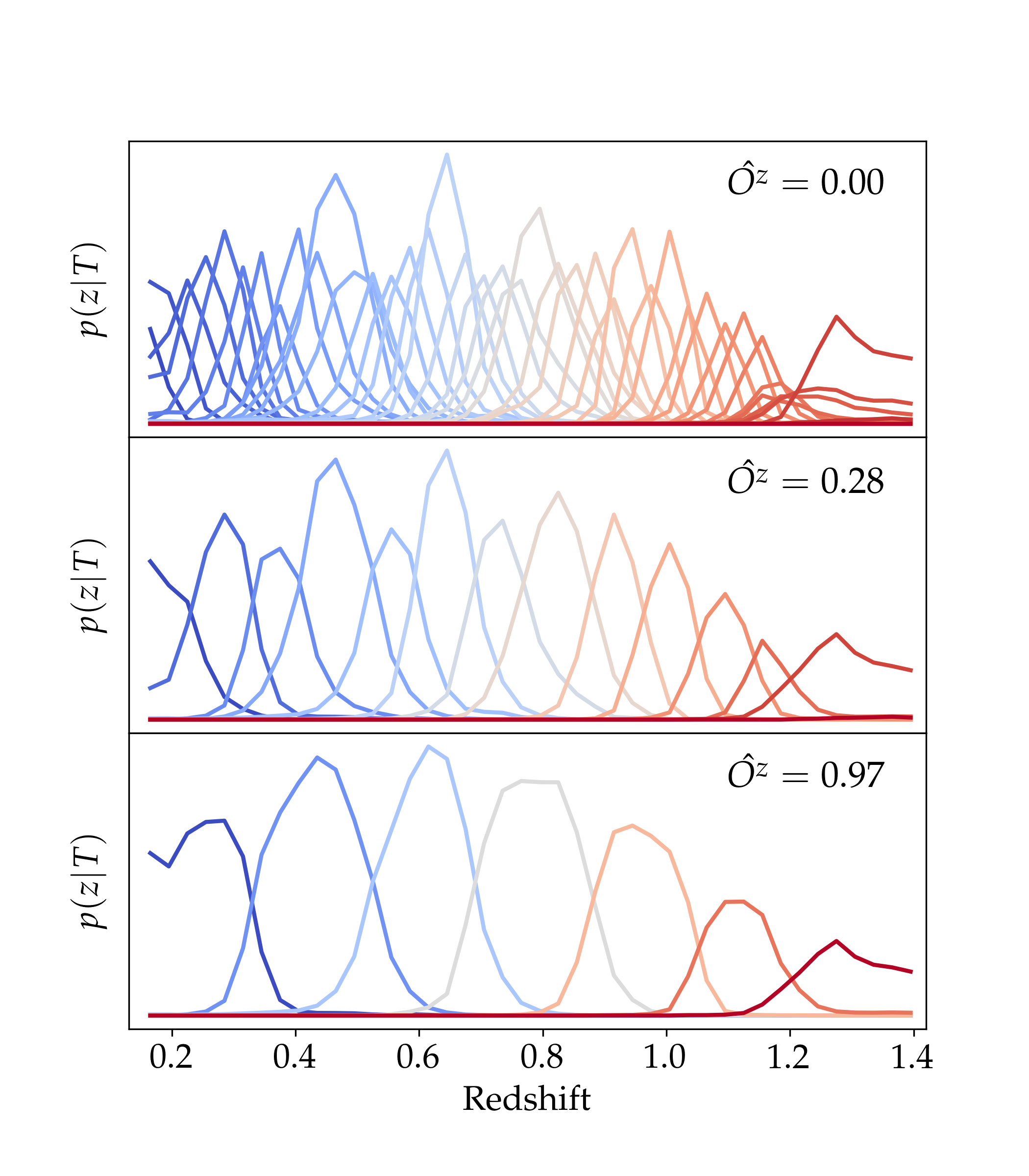}
	\caption{Redshift distributions for different superphenotype prescriptions used in Figure \ref{fig:corrs}. The superphenotypes are constructed by joining phenotypes that share one, three and six average redshift values (top to bottom), and we show the corresponding values of the overlap metric of Equation (\ref{eq:overlap}). As expected, superphenotypes that show small redshift overlap between them have a metric values close to unity, while overlaping ones show a smaller value.}
\label{fig:nzTs}
\end{figure}

Figure \ref{fig:uncertainties} shows the shot noise and sample variance redshift uncertainties for the simulated calibration samples described at the beginning of this Section: \textit{healpixel} samples are shown in red and \textit{random-equivalent} samples in blue. While the random-equivalent samples only contain shot noise, the healpixel samples also include the contribution from sample variance. In the upper-left panel, we show violin plots with the distribution of $N_z$'s for each redshift bin.  In the upper-right panel we show the normalized variance, Equation (\ref{param2}), as a function of redshift for the two sets. As it is clear from the plot, the case with sample variance shows a $>10\times$ larger normalized variance compared to shot noise. The shot noise contribution to the normalized variance shows a flat behavior in redshift and a steady value of the normalized variance around 1, as expected for a Poisson distribution. Related to this, the lower-right panel of Figure \ref{fig:uncertainties} shows the contribution from sample variance alone, as parametrized in Equation (\ref{param2}) by $\mathrm{Var}(\Delta_z)$. For this term, the shot noise case is consistent with zero, as expected, and the sample variance case shows a decreasing trend with redshift, as expected for equally-sized redshift bins.

From these two different sources of uncertainty, shot noise and sample variance, and the corresponding sets of redshift distributions derived from them, we can compare the uncertainty they introduce in recovering the mean redshift of the population, as this is a very important quantity for weak lensing analyses.  We calculate the mean $z$ for each simulated spectroscopic survey, and
the lower-left panel of Figure \ref{fig:uncertainties} plots the histogram of mean $z$'s for the members of the healpixel and random-equivalent samples.
The healpixel set, which includes sample variance, presents a scatter on the mean redshift of the distribution that is $\approx10\times$ larger than that of the random-equivalent sample, which has only shot noise.

\subsection{Effects in phenotype}
\label{sec:phenotypes}

\begin{comment}
\begin{figure}
\centering
	\includegraphics[width=0.5\textwidth]{figure/vart.png}
	\caption{Median normalized variance in the superphenotypes as introduced by sample variance, as a function of the number of redshift bins used to bin superphenotypes (the 0 in that plot shows the variance when not binning phenotypes together. For characterization of this uncertainty, we denote $\bar{\lambda_T}$ to the median normalized variance for the superphenotype definition that joins phenotypes in groups of 6 contiguous redshift bins. )}
\label{fig:vart}
\end{figure}
\end{comment}

The left panel of Figure~\ref{fig:corrs} presents the correlation matrix of $N_t$, the number of spectroscopic galaxies per phenotype $t$ in a spectroscopic survey, for the healpixel simulation set.
We observe a strong pattern of correlations between phenotypes, which must be caused by
sample variance (\ie~coming from different patches in the sky), since shot noise will induce no inter-phenotype correlations.
Sample variance will produce physical clustering of galaxies in redshift, but due to the intrinsic correlation between phenotype and redshift, it will also result in a correlation among different phenotypes that live at similar redshifts. 

Since we know that types being at similar redshifts is the cause of their correlation induced by sample variance, we segment the phenotypes into groups with similar redshifts, which we will call \textit{superphenotypes} ($T$).  We do this by assigning each phenotype $t$ to one of the 42 redshift bins $z$ (as defined in \S\ref{sec:sims}) according to the mean of the $p(z|t)$ as determined by spectroscopy.
In the right half of Figure \ref{fig:corrs} we show the correlation matrices of superphenotypes $T$ that are constructed by joining all phenotypes $t$ that are assigned to groups of one, two, three, or six redshift bins (left to right, top to bottom in the Figure). That is, in the first case we have as many superphenotypes as redshift bins, in the second case we have half that amount, and so on (as can be seen by the dimension of the correlation matrices).  As expected, the correlation matrices look much more diagonal than that in the left half of the Figure.

In order to choose a prescription for joining phenotypes to make superphenotypes that are disjoint in redshift, and hence independent under sample variance, we devise the following metric $\hat{O^z}$. For a given superphenotype definition $T$, \ie~a number of phenotypes with the same mean redshift that are joined into one superphenotype, we estimate the corresponding redshift overlap matrix $O^z_{ij}$ from their redshift distribution $p(z,T)$ as:
\begin{equation}
O^z_{ij} = \frac{\sum_z p(z|T_i) p(z|T_j)}{\sqrt{\left(\sum_z p(z|T_i)^2\right)\left(\sum_z p(z|T_j)^2\right)}}
\end{equation}
and compute its determinant corrected by a dimensional factor: 
\begin{equation}
\hat{O^z} = \mathrm{Det} (O^z_{ij})^{N^T/N^z}.
	\label{eq:overlap}
\end{equation}

This metric $\hat{O^z}$, defined to be between 0 and 1, will become closer to unity if the superphenotypes present small overlap in redshift, and will be smaller otherwise. Figure \ref{fig:nzTs} shows the redshift distributions of superphenotypes in three different prescriptions (joining one, three and six redshift bins), and the corresponding values of the metric. We can see how the metric becomes close to unity for the case where superphenotypes are well separated in redshift. Later on, we will also use this mechanism for devising a sampling method that includes sample variance uncertainties in Section \ref{sec:method}.  

%%%%%%%%%%%%%%%%%%%%%%%%%%%%%%%%%%%%%%%%%%%%%%%%%%%%%%%%%%%%%%%%%%%%%%%%%%%%%%%%%%%%%%%%%%%%%%%%%%%%%%%%%%%%%%%%%%%%%%%%
\section{Theoretical estimate of sample variance}
\label{sec:theory}
%%%%%%%%%%%%%%%%%%%%%%%%%%%%%%%%%%%%%%%%%%%%%%%%%%%%%%%%%%%%%%%%%%%%%%%%%%%%%%%%%%%%%%%%%%%%%%%%%%%%%%%%%%%%%%%%%%%%%%%%

In the last section we characterized the effects of sample variance in redshift priors from calibration samples. 
Equation (\ref{param2}) describes that variance in the redshift distribution of sky patches given the galaxy density in them, an estimate of the shape of the distribution, and an estimate of the sample variance contribution, independent of galaxy density. 
The first two ingredients can be obtained directly from the data, while the sample variance term, which can be the leading contribution to the redshift uncertainty of calibration patches, cannot be obtained from the data itself, and has only been estimated with simulations so far. 

In this section we discuss how to compute this sample variance contribution for a given cosmological model.

We want to compute the variance of number density fluctuations over the entire calibrator sample that we assume covers a fraction of sky $f_{\rm sky} = A/ 4\pi$ given its area of $A$.
To simplify the calculation we assume that the calibrator survey geometry is circular so that the angular scale of the survey is related to its area by
$A_f = 2\pi (1-\cos(\theta_f))$.
This approximation works well in practice and the results of this section can be straightforwardly extended to more complicated survey geometries if necessary.

The galaxy fluctuation field, coarse grained over an angular scale $\theta_f$, is the spherical convolution of the fluctuation field with an angular smoothing filter $W$:
\begin{align}
\hat{\Delta}_z (\vec{n}) = \int W(\vec{n}\cdot\vec{n}') \, \Delta_z(\vec{n}') \, d\Omega' \,,
\end{align}
that we assumed depends only on angular separation $\cos(\theta) = \vec{n}\cdot\vec{n}'$ between two points.

In this case, similarly to the flat case, the convolution of two functions is equivalent to the product of their Fourier transforms.
If we expand the smoothing filter in a Fourier-Legendre series $W(x) = \sum_\ell \tilde{W}_\ell P_\ell(x)$, in terms of Legendre polynomials $P_\ell$, it can be shown that the smoothed density field can be decomposed in spherical harmonics as:
\begin{align}
\hat{\Delta}_z(\vec{n}) = \sum_\ell \frac{4\pi}{2\ell+1} \tilde{W}_\ell \sum_{m=-\ell}^\ell a_{\ell m} Y_{\ell m}(\vec{n}) \,,
\end{align}
where $Y_{\ell m}$ are spherical harmonics with coefficients $a_{\ell m}$.
Since the smoothing filter depends only on angular separation we can center the reference vector at the north pole $\vec{p}$ and compute the coefficients of the Fourier-Legendre expansion as:
\begin{align}
\tilde{W}_\ell = \sqrt{\frac{2\ell+1}{4\pi}} \int W(\vec{p} \cdot \vec{n}) \, Y_{\ell 0}(\vec{n}) \, d\Omega \,.
\end{align}

Here we are interested in an angular top-hat smoothing filter, $W(x) \propto \Theta(\theta_f-{\rm arcos}(x))$ where $\theta_f$ is the angular aperture of the top-hat. 
The Fourier-Legendre coefficients can be computed and result in:
\begin{align}
F_\ell \equiv \frac{4\pi}{2\ell+1}\tilde{W}_\ell = \frac{2\pi}{A_f}\frac{P_{\ell-1}(\cos\theta_f)-P_{\ell+1}(\cos\theta_f)}{2\ell+1} \,.
\end{align}

With these we can compute the angular correlation function of the smoothed galaxy density field at two different redshifts $z$ and $z'$:
\begin{align}
\langle \hat{\Delta}_z(\vec{n} )\,\hat{\Delta}_{z'}(\vec{n}') \rangle = \sum_\ell \frac{2\ell+1}{4\pi} \, P_\ell( \vec{n}\cdot\vec{n}' ) \, F_\ell^2 \mathcal{C}_\ell^{zz'} \,,
\end{align}
which is very similar to the standard result for the correlation function in terms of the harmonic power spectrum, $\mathcal{C}_\ell$, of galaxy number counts fluctuations with the difference that multipoles are here weighted differently because of the smoothing filter.

The covariance of the smoothed filter is given by:
\begin{align} \label{Eq:TheorySV}
{\rm SV}(z,z') \equiv {\rm Cov} (\hat{\Delta}_z,\hat{\Delta}_{z'}) = \sum_\ell \frac{2\ell+1}{4\pi} \, F_\ell^2 \mathcal{C}_\ell^{zz'} \,,
\end{align}
which is the equation that we use to compute the sample variance term once the harmonic power spectra are computed for a given cosmological model.
For the calculation of the theory galaxy number counts power spectrum we use CAMB~\cite{Lewis:1999bs} and follow the discussion in~\cite{Challinor:2011bk} for the different effects to include in the calculation.
We treat the modeling of bias and its redshift dependence separately, as discussed in the next sections, and assume that it is scale independent.

% Discussion of the mask from the simulation.
The simulation from which we extract the sample variance measurement that we seek to match is run with a finite volume.
This means that part of the sky is masked and we need to include this effect in the theory calculation.
The power spectrum for the masked fluctuations, $\tilde{\mathcal{C}}_\ell$, is computed as $\tilde{\mathcal{C}}_\ell = \sum_{\ell'} M_{\ell \ell'}\mathcal{C}_{\ell'}$~\citep{Hivon:2001jp} where the mode-coupling matrix, $M_{\ell \ell'}$ is computed with~\cite{Alonso:2018jzx}.

%%%%%%%%%%%%%%%%%%%%%%%%%%%%%%%%%%%%%%%%%%%%%%%%%%%%%%%%%%%%%%%%%%%%%%%%%%%%%%%%%%%%%%%%%%%%%%%%%%%%%%%%%%%%%%%%%%%%%%%%
\subsection{Comparison with simulation} \label{SubSec:SimulationComparison}
%%%%%%%%%%%%%%%%%%%%%%%%%%%%%%%%%%%%%%%%%%%%%%%%%%%%%%%%%%%%%%%%%%%%%%%%%%%%%%%%%%%%%%%%%%%%%%%%%%%%%%%%%%%%%%%%%%%%%%%%

From the MICE2 simulation, discussed in Section~\ref{sec:sims}, we can extract redshifts and positions of galaxies in their rest frame, hence neglecting line-of-sight effects distorting both.
In this situation sample variance is only sourced by CDM fluctuations.

We start the comparison by considering the ratios of sample variance for different scales of the calibrator survey.
With the approximation that bias between CDM and galaxy number-count fluctuations is scale independent, these ratios do not depend on the modeling of bias and in particular do not depend on its redshift evolution.

We consider four areas for the calibrator survey: 13.4 deg$^2$, 3.36 deg$^2$, 0.893 deg$^2$ and 0.21 deg$^2$, which corresponds to healpix nside values of 16, 32, 64 and 128, respectively.

Sample variance estimated from the simulations is expected to be noisy and we can compute the expected covariance of sample variance as:
\begin{align} \label{Eq:SVvariance}
& {\rm Cov}( {\rm SV}(z_1,z_2), {\rm SV}(z_3,z_4) ) = \nonumber \\
& \hspace{0.5cm} = \sum_{\ell_1,\ell_2} \frac{(2\ell_1+1)(2\ell_2+1)}{(4\pi)^2} F_{\ell_1}^2 F_{\ell_2}^2 \, {\rm Cov}(C_{\ell_1}^{z_1 z_2},C_{\ell_2}^{z_3 z_4}) \,.
\end{align}
In the following we neglect, for simplicity, any non-Gaussian 
component to the power spectrum covariance that is then given by:
\begin{align} \label{Eq:ClsCovariance}
{\rm Cov}(C_{\ell_1}^{z_1 z_2},C_{\ell_2}^{z_3 z_4}) = \delta_{\ell_1,\ell_2} \frac{\mathcal{C}_{\ell_1}^{z_1 z_3}\mathcal{C}_{\ell_1}^{z_2 z_4}+\mathcal{C}_{\ell_1}^{z_1 z_4}\mathcal{C}_{\ell_1}^{z_2 z_3}}{f_{\rm sky}(2\ell_1+1)} \,,
\end{align}
where $f_{\rm sky}$ is the sky fraction covered by the simulation.

\begin{figure}
\centering
\includegraphics[width=\columnwidth]{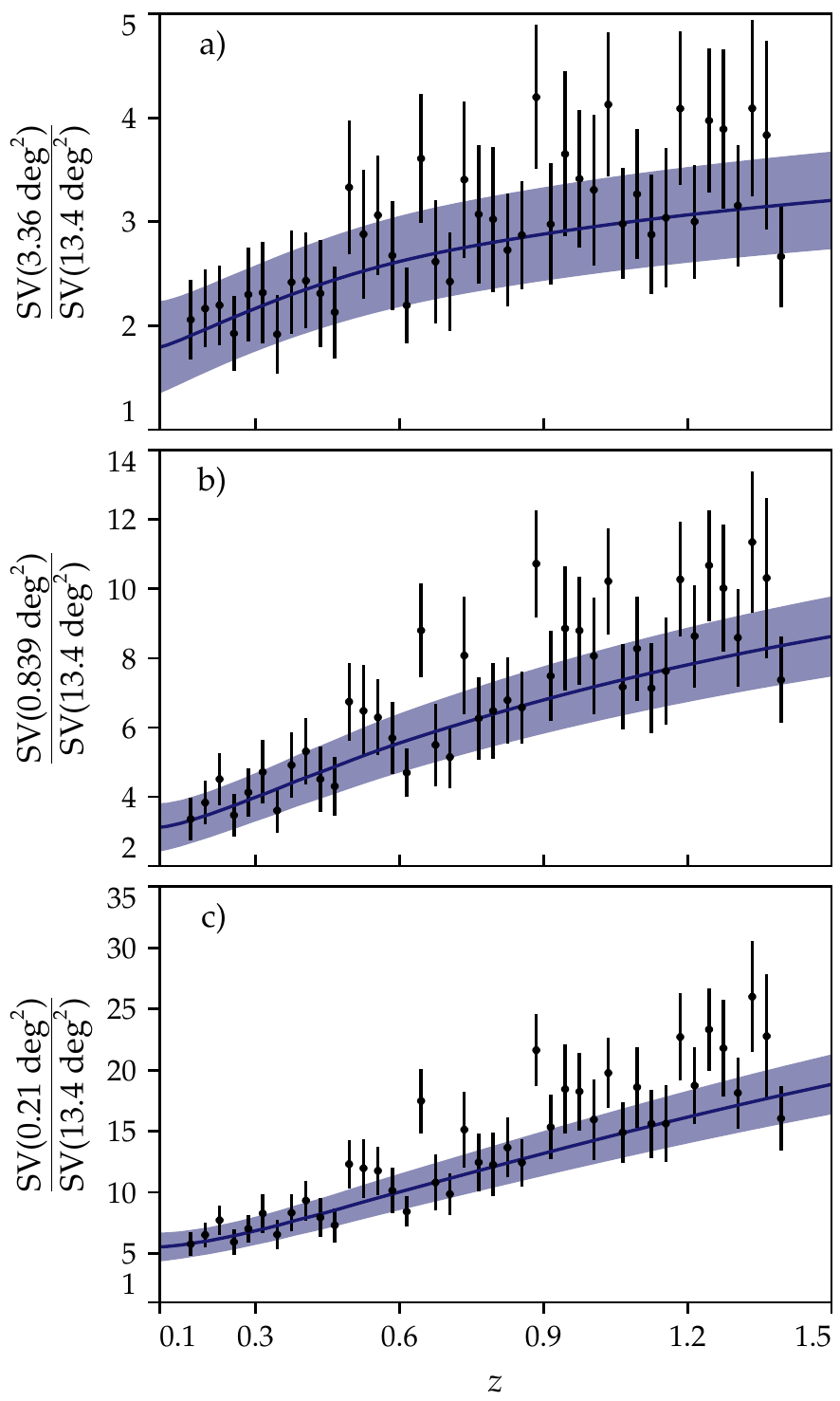}
\caption{ \label{Fig:SVTheoryRatio}
Ratio of sample variance at different angular scales.
Theory predictions (continuous line) are computed with Eq.~\ref{Eq:TheorySV} and their variance is computed with Eq.~\ref{Eq:SVvariance}.
Measurements from simulation are shown as dots with bootstrap error bars, as discussed in Sec.~\ref{sec:sims}.
Notice that the results shown in this plot are independent of the modeling of bias.
Moreover, since cosmological parameters of the simulation are known, there is no free parameter to optimize to obtain these results.
}
\end{figure}

In Figure~\ref{Fig:SVTheoryRatio} we show the ratio of sample variances with respect to sample variance at the largest scale.
As we can see all panels show agreement with the theoretical prediction (\ref{Eq:TheorySV}), within the error bars, at the $10\%-15\%$ level.
Notice that there are no free parameters to obtain these results since the cosmology in the simulation is known and (scale-free) galaxy bias cancels in the ratio.

Some of the discrepancy between the theory prediction and the simulation measurement in Figure~\ref{Fig:SVTheoryRatio}, especially at high redshift, is due to the approximation that we have made about survey geometry.
To test this we have computed the sample variance term with Eq.~\ref{Eq:TheorySV} but using the harmonic power spectra measured from the simulation finding better agreement.

To compute the full theory prediction of the sample variance we need to model bias and its redshift dependence (one bias value per redshift bin), between galaxies and CDM. This step is crucial to get a reliable estimate of sample variance and hence we devise two ways of doing it that are complementary and can be used to check the reliability of the theoretical prediction.

The first method consists in computing bias from the halo model (see~\cite{Cooray:2002dia} for a review). 
Since we do not have very strict accuracy requirements we model the halo mass function following~\cite{Sheth:1999mn}.
To connect this to galaxy counts for our magnitude limited simulation we need to model the conditional luminosity function that we coarsely approximate by assuming that galaxy luminosity is proportional to halo mass.
In this case the halo model gives a prediction for the number of galaxies as a function of redshift that depends on two parameters: the proportionality constant between halo mass and luminosity and the effective mass cut of the simulation $M^*$.
We fit the simulation $\bar{N}(z)$ to get the best estimate for these two parameters and compute bias as the logarithmic derivative of the $\bar{N}(z)$ as a function of the mass cutoff.

The second method consists in measuring bias within the calibration survey. To do so we compute the correlation function of galaxies within the small patch and compare it with the theory prediction, after subtracting shot noise from the measured correlation function.
If the theory power spectrum is computed with unit bias and assuming that the measured $a_{\ell m}$ are Gaussian distributed then the estimate of bias at each redshift $z_i$ is given by:
\begin{align} \label{Eq:BiasFromSims}
b^2(z_i) = \frac{\sum_{\ell}(2\ell+1)\, \mathcal{C}_\ell^{z_i z_i\, {\rm obs}}/\mathcal{C}_\ell^{z_i z_i}}{\sum_{\ell}(2\ell +1)} \,,
\end{align}
where, for simplicity, we have neglected the correlation between the power spectra at different redshifts that is negligible for non-neighboring 
redshift bins.

The error on the bias determination can be easily computed from the covariance of the observed power spectra in Eq.~\ref{Eq:ClsCovariance}.

Comparing the two approaches we find that they agree at the $10\%$ level which is sufficient for our application. 
Better agreement can be likely achieved by refining the theoretical modeling of bias.

\begin{figure*}
\centering
\includegraphics[width=\textwidth]{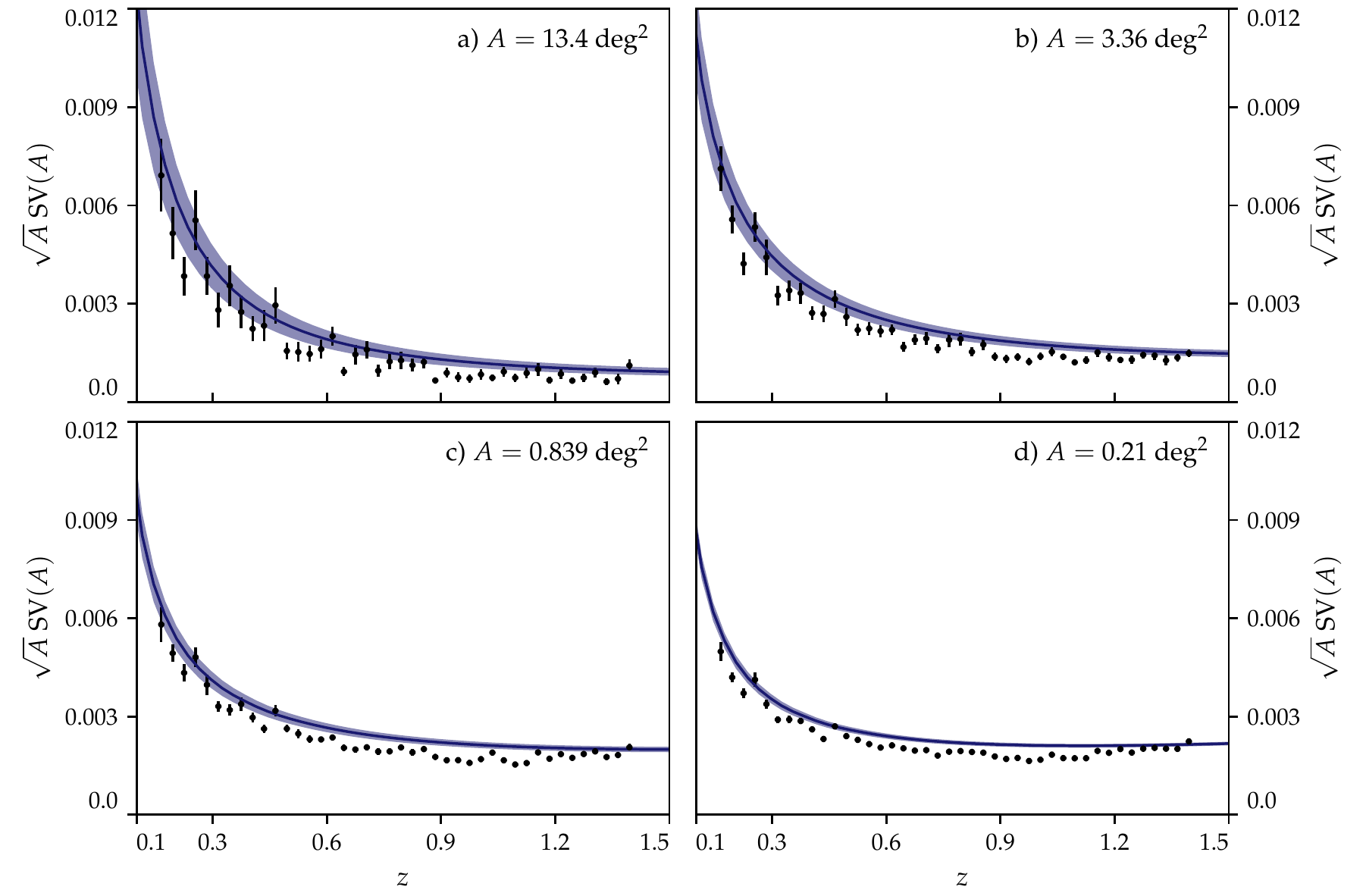}
\caption{ \label{Fig:SVFullTheory}
Sample variance at different angular scales for the MICE simulation.
Theory predictions (continuous line) are computed with Eq.~\ref{Eq:TheorySV} and their variance is computed with Eq.~\ref{Eq:SVvariance}.
Measurements from simulation are shown as dots with bootstrap error bars, as discussed in Sec.~\ref{sec:sims}.
}
\end{figure*}

With a model for bias we can compute the complete theory prediction for sample variance which, for the four angular sizes that we consider, is shown in Figure~\ref{Fig:SVFullTheory}.
As we can see the agreement between the measurement and theoretical prediction is good at the $10\%-20\%$ level.
The theory line slightly overestimates the measurement mostly due to the modeling of bias.

%%%%%%%%%%%%%%%%%%%%%%%%%%%%%%%%%%%%%%%%%%%%%%%%%%%%%%%%%%%%%%%%%%%%%%%%%%%%%%%%%%%%%%%%%%%%%%%%%%%%%%%%%%%%%%%%%%%%%%%%
\subsection{Dependence on finite simulation volume} \label{SubSec:FiniteVolume}
%%%%%%%%%%%%%%%%%%%%%%%%%%%%%%%%%%%%%%%%%%%%%%%%%%%%%%%%%%%%%%%%%%%%%%%%%%%%%%%%%%%%%%%%%%%%%%%%%%%%%%%%%%%%%%%%%%%%%%%%

With the theory prediction for sample variance at hand we can test the impact of fluctuations above the scale of the MICE simulation.
To do so we need to compute the theory sample variance neglecting the impact of the simulation mask and the way in which it suppresses power.

\begin{figure}
\centering
\includegraphics[width=\columnwidth]{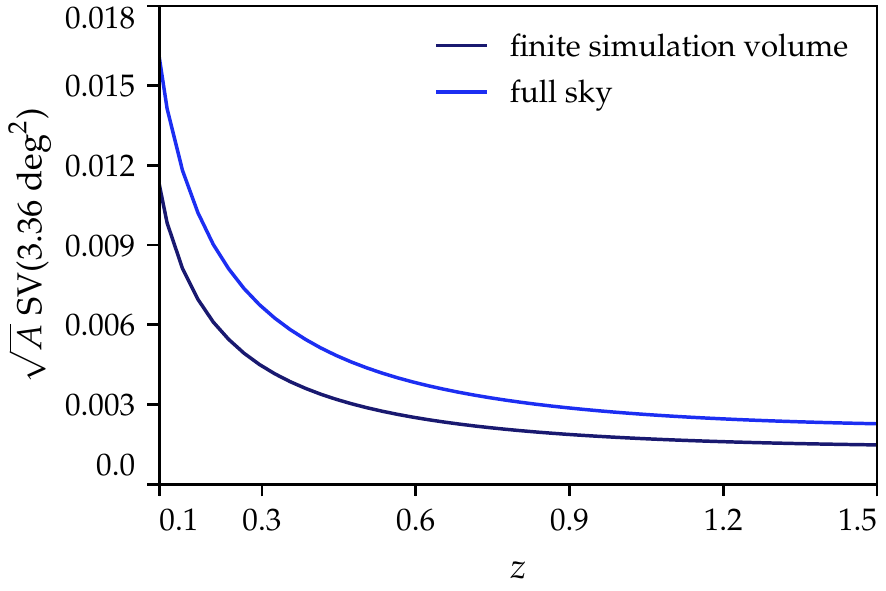}
\caption{ \label{Fig:SVFiniteVolume}
Sample variance computed with and without the contribution of modes larger than the MICE simulation box, as shown in legend.
}
\end{figure}

The result is shown in Figure~\ref{Fig:SVFiniteVolume} along with the previous result for comparison, for one of the calibrator survey scale in particular.
As we can see the effect of these large scale modes can be significant and in particular the estimate of sample variance from the simulation, in this case, underestimates it by about $30\%$.
It is therefore advisable to use the theoretical prediction than the simulation values for sample variance when analyzing real data.

%%%%%%%%%%%%%%%%%%%%%%%%%%%%%%%%%%%%%%%%%%%%%%%%%%%%%%%%%%%%%%%%%%%%%%%%%%%%%%%%%%%%%%%%%%%%%%%%%%%%%%%%%%%%%%%%%%%%%%%%
\subsection{Dependence on cosmological model} \label{SubSec:SVCosmologyDependence}
%%%%%%%%%%%%%%%%%%%%%%%%%%%%%%%%%%%%%%%%%%%%%%%%%%%%%%%%%%%%%%%%%%%%%%%%%%%%%%%%%%%%%%%%%%%%%%%%%%%%%%%%%%%%%%%%%%%%%%%%

\begin{figure}
\centering
\includegraphics[width=\columnwidth]{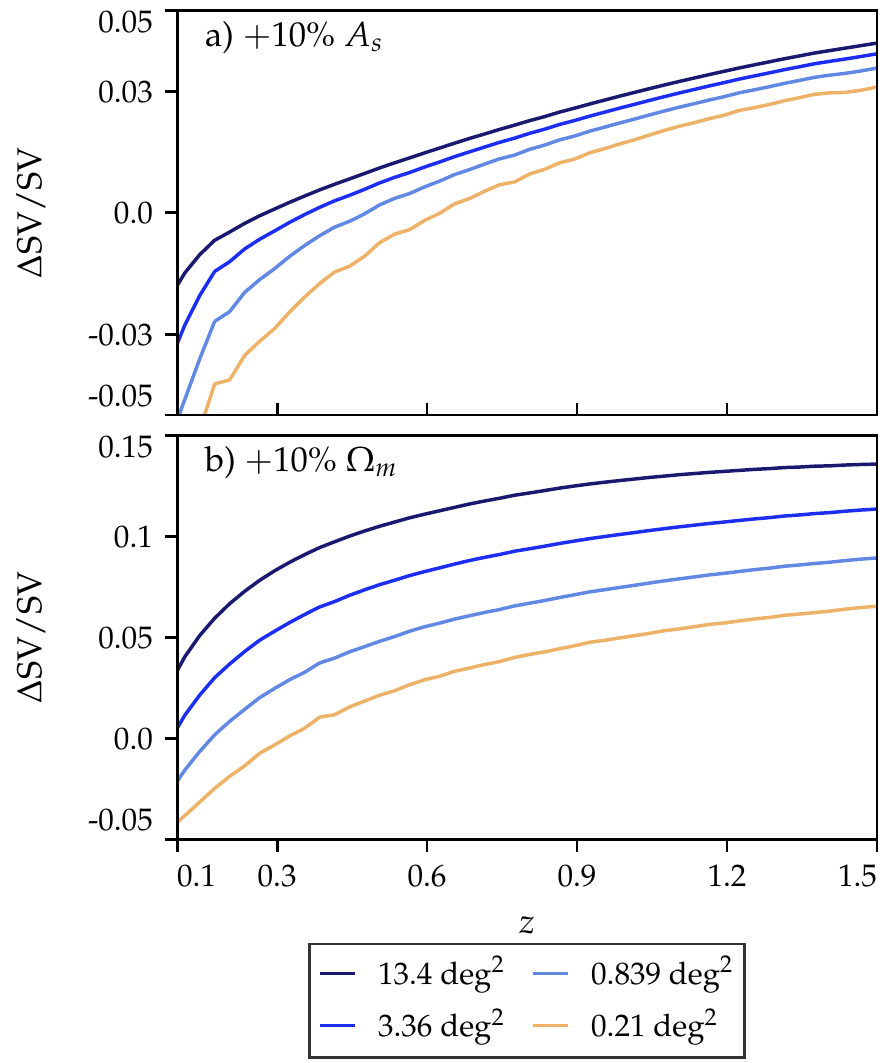}
\caption{ \label{Fig:SVCosmologyDependence}
Relative differences in sample variance estimates as a result of variations of cosmological parameters.
Different lines show different angular scales, as shown in legend.
}
\end{figure}

In this section we discuss the dependence of sample variance on cosmological parameters.
We test the impact of all relevant $\Lambda$CDM parameters: matter density $\Omega_m$, the amplitude and tilt of the primordial scalar power spectrum, $A_s$ and $n_s$ respectively, and the Hubble constant $H_0$.
We find that $\Omega_m$ and $A_s$ influence it the most, as shown in Fig.~\ref{Fig:SVCosmologyDependence}.
Note that a variation in $A_s$, in Fig.~\ref{Fig:SVCosmologyDependence}a), is not simply rescaling the entire theory prediction.
This happens because bias is not kept fixed. We fix the halo model parameters that produce the bias prediction that would then take into account the variation of the number of observed objects when varying cosmological parameters.
A similar effect happens for a variation in $\Omega_m$.
%

%%%%%%%%%%%%%%%%%%%%%%%%%%%%%%%%%%%%%%%%%%%%%%%%%%%%%%%%%%%%%%%%%%%%%%%%%%%%%%%%%%%%%%%%%%%%%%%%%%%%%%%%%%%%%%%%%%%%%%%%
\subsection{Other effects} \label{SubSec:SVOtherEffects}
%%%%%%%%%%%%%%%%%%%%%%%%%%%%%%%%%%%%%%%%%%%%%%%%%%%%%%%%%%%%%%%%%%%%%%%%%%%%%%%%%%%%%%%%%%%%%%%%%%%%%%%%%%%%%%%%%%%%%%%%

In this section we discuss the dependence of sample variance on other modeling assumptions that we have made.

\begin{figure}
\centering
\includegraphics[width=\columnwidth]{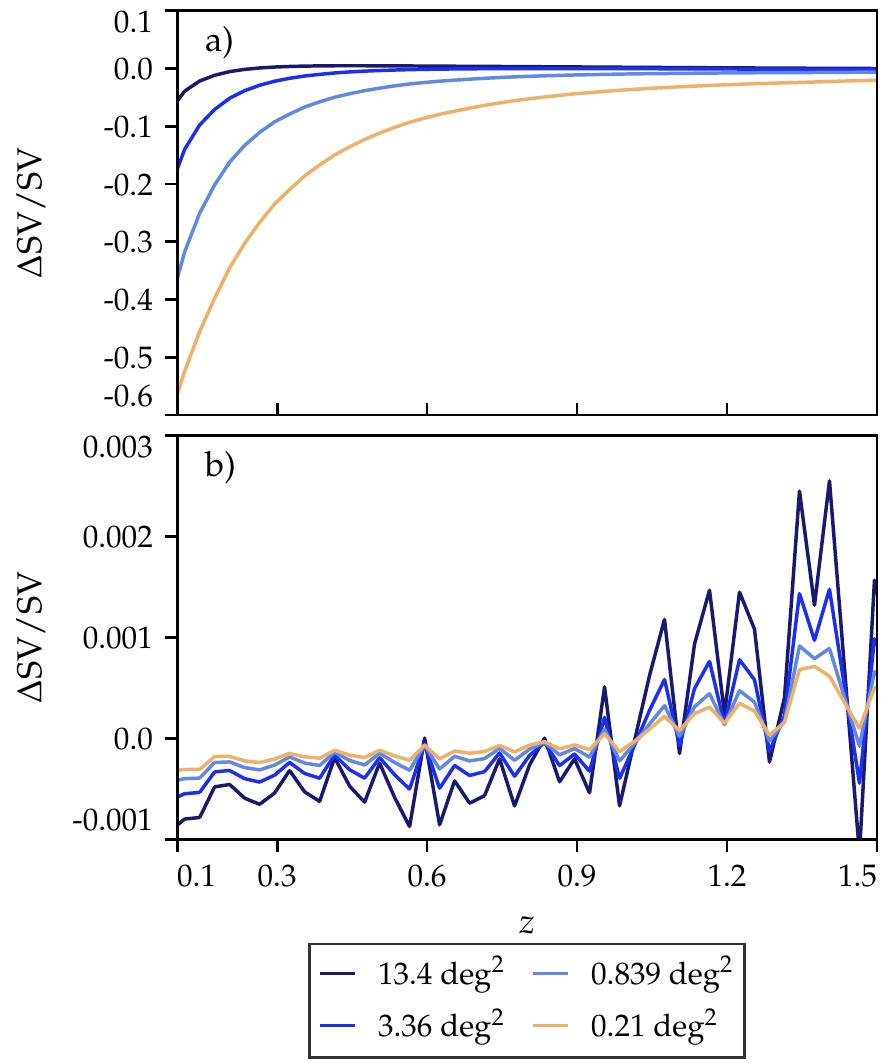}
\caption{ \label{Fig:SVOtherEffects}
Relative differences in sample variance estimates using the linear and non-linear matter power spectrum, Panel a), and magnification, redshift space distortion and other relativistic effects along the observer line of sight, Panel b).
Different lines show different angular scales, as shown in legend.
}
\end{figure}

First of all we consider the impact of non-linear growth of matter perturbations.
We show in Panel a) of Fig.~\ref{Fig:SVOtherEffects} the difference in the SV prediction when using the non-linear modeling in~\cite{Takahashi:2012em} and only linear perturbation theory.
As we can see SV in linear theory would be in general smaller and the difference between the two theory predictions increases as redshift decreases as we would expect. 
As expected this discrepancy strongly depends on the size of the calibrator patch.
For small areas the effect is large and between 50\% and 10\% for the redshifts that we consider. 
At the two largest scales instead the effect is very small and in general sub percent.

Then we consider other effects that were neglected in the comparison with simulations in the previous section.
These include all relativistic corrections to galaxy number counts fluctuations, discussed in~\cite{Challinor:2011bk}.
In particular these include lensing magnification and redshift space distortions.
Magnification bias in this case can be easily extracted from the MICE simulation and included in the calculation of the theory power spectra.
As we can see in Panel b) of Fig.~\ref{Fig:SVOtherEffects} the combination of these effects is negligible for our purposes.
Note that the theory prediction is noisy because magnification bias extracted from the simulation is noisy.

\section{Sampling over a redshift prior}
\label{sec:method}

In the previous sections we have studied shot noise and sample variance as sources of uncertainty in the characterization of a redshift distribution coming from observations in a limited area in the sky. We have found sample variance to be the dominant source of uncertainty, and we have characterized its effects in redshift priors using simulated calibration samples and theoretical estimation. In this section we will present a method to sample the redshift information from a calibration survey that manifests the proper level of sample variance. In the next two sections we will validate this method in simulations and apply it to data.

We will make extensive use of the Dirichlet distribution for sampling our priors. The Dirichlet distribution is a family of continuous multivariate probability distributions parametrized by a vector of positive reals, or integers, in our case. They are commonly used as prior distributions in Bayesian statistics, exploiting the fact that the Dirichlet distribution is the conjugate prior of the multinomial distribution. This means that when we have a multinomial likelihood and choose the prior to be Dirichlet distributed, the posterior will also be Dirichlet distributed.  This makes the posterior sampling straightforward. 

In a general example where $N$ is a vector with elements $n_i$, $i=1,\ldots,M$, the corresponding fractions $f_i$ given the $n_i$ values are Dirichlet distributed as:

\begin{equation}
	\mathrm{Dir}(\{f_i\};\{n_i\}) \propto \prod_{i=1}^{M}f_{i}^{n_{i}-1}.
\end{equation}

$N$ can, for example, be the redshift distribution of a galaxy sample coming from a patch in the sky, and then we can say the prior probabilities on $p(z) = f_z$ from that patch follow a Dirichlet distribution $\mathrm{Dir}(N)$. Conveniently, the Dirichlet sampling will yield samples of $f_z$ that fulfill two required properties of redshift probabilities, namely $f_z > 0$ $\forall z$ and $\sum_z f_z = 1$. 

It is also important to note that the Dirichlet model does not carry any assumption on the nature of the bins of the input vector, so the bins can be of an arbitrary width in $z$ or have different physical interpretations---they are just different categories in our model. Nonetheless, once such categories are defined, for instance in the prior term, they should be kept consistent for the estimation of the posterior.

Two properties of the Dirichlet distribution will be of particular use: first, $\langle f_i \rangle = n_i / n_{\rm tot},$ with $n_{\rm tot}\equiv\sum_j n_j,$ such that the expectation value is equal to the distribution of the sample set.  Second, the variance of  $f_i$ is approximately $n_i / n_{\rm tot}^2$.
The latter means that 
if we rescale all of the $n_i \rightarrow n_i / \lambda,$ then the mean of the Dirichlet distribution is unchanged, but $\textrm{Var} f_i \rightarrow \lambda \, \textrm{Var} f_i.$  We will make use of this property below as a means to inflate the shot-noise variation intrinsic to the Dirichlet distribution so that it approximates sample variance plus shot noise.

\subsection{Bootstrap sampling (BT)}

We can produce samples of a prior $p(z,t)$ distribution by using the bootstrap resampling scheme on a redshift sample with known $z,t$ pairs. We can produce an arbitrary number of calibration samples which are just random resamplings, with repetition, of the original one. Bootstrap resampling will treat galaxies at all redshifts in the same way, and will not resample the sample variance. We will use the lable ``BT'' for bootstrap resampling.

\subsection{Basic Dirichlet sampling (Dir)}
\label{sec:basic_dir}

In the past \citep{Sanchez2019,Alarcon2019}, we have drawn samples from the prior using a Dirichlet distribution on the counts $M_{zt}$ of calibration galaxies in joint bins of redshift and phenotype. The Dirichlet distribution treats the 2d array of counts as a single 1d array of categories, so that:

\begin{equation}
	\mathrm{Dir}(\{f_{zt}\};\{M_{zt}\}) \propto \prod_{z=1}^{N_z}\prod_{t=1}^{N_t}f_{zt}^{M_{zt}-1}.
\end{equation}

When doing this, we are assuming all redshifts and phenotypes have the same uncertainties, uncorrelated beyond the constraint $\sum f_{tz}=1,$ \ie\ we are not considering sample variance.
This ``Dir'' case assumes that the prior information on redshift and on phenotype both arise from the same calibration sample.
We might, however, often have stronger prior information on the phenotype distribution compared to the redshift distribution, \ie\ we have more galaxies with definitive phenotypes (using high quality photometry) than have definitive redshifts (using spectroscopic or many-band photometric information). 

\subsection{3 step Dirichlet sampling (3sDir)}
\label{sec:3sDir}

Now we will construct a sampling method based on the Dirichlet model but using the characterization of sample variance from Section \ref{sec:uncertainties}. Instead of considering the redshift and phenotype parts together as in \S\ref{sec:basic_dir}, now we will split the problem by making use of the following relations: 
\begin{align}
  p(z,t) & = p(t | z, T) p(z | T) p(T) \\
  f_{zt} & =f_{t}^{zT} f_z^T f_T
\end{align}
where the mean of each fraction $f$ in the second row is given by its corresponding term in the first row.  All $f$'s must be non-negative, and there are sum constraints $\sum_{zt} f_{zt} = \sum_t f_t^{zT} = \sum_z f_z^T = \sum_T f_T=1.$

The 3-step Dirichlet sampling method consists of drawing, in sequence, values of $f_T$, then $f_z^T$, then $f_t^{zT}$ from individual Dirichlet distributions, using appropriate source counts from the calibration survey to define the $n_i$ in each.  When sampling distributions that span multiple redshifts (namely $f_z^T$ and $f_T$) we will however rescale the $n_i$ so as to inflate the variance of the Dirichlet distribution to the level expected for the sum of shot noise and sample variance.  
The key input is the normalized variance from Equation~(\ref{param2}) which we will label as  $\lambda_z \equiv \mathrm{Var}(N_z)/N_z$.
The terms on the right-hand side of the equation can be estimated from theory as in Section \ref{sec:theory}.  

The first sampling step is to draw a set of $f^T$ given the counts $M_T$ of each superphenotype in the calibration sample.  For this purpose we define $\bar\lambda = \langle \lambda_z \rangle $ as the mean ratio of (shot$+$sample) variance to shot noise.  We then draw $f^T$ from a Dirichlet distribution that scales the observed counts in a manner that generates the desired total noise:
\begin{enumerate}[leftmargin=1cm]
	\item $f_T \leftarrow \mathrm{Dir}(M_T/\bar\lambda)$
  \setcounter{enumCount}{\value{enumi}}
\end{enumerate}
Recall that the superphenotypes were chosen to span large enough redshift ranges that their counts are nearly uncorrelated by sample variance.  Thus this noise-inflated Dirichlet draw will approximate the conjugate to the true distribution for $M_T,$ which is essentially a coarse redshift binning of the sources.  The method is not exact, in the sense that the real $\lambda_z$ do change with $z$ (Figure~\ref{fig:uncertainties}, upper right), while our Dirichlet partition must assume a fixed $\bar\lambda$ over the full range.

One important facet of this stage is that it does not require redshift information.  The counts $M_T$ can be made over all calibration fields with deep photometry sufficient to assign (super)-phenotypes, without the need for redshift assignments.  This allows reduction of shot noise and sample variance in this step---which was noted by \citet{Buchs2019} as the largest noise source in their implementation of phenotypic redshifts.
        
The next step is to draw values of $f_z^T$, \ie\ distribute the probabilities $f_T$ into redshift bins.
We compute for every superphenotype $T$  the noise excess $\lambda_T = \sum_z \lambda_z \: p(z|T) \approx \sum_z \lambda_z M_{zT}/M_T$.  A sample of the $f_{zT}=f_z^Tf_T$ are then generated by
\begin{enumerate}[leftmargin=1cm]
  \setcounter{enumi}{\value{enumCount}}
	\item $f_{zT} \leftarrow \mathrm{Dir}(M_{zT}/\lambda_T) \: f_T$
  \setcounter{enumCount}{\value{enumi}}
\end{enumerate}
Here $M_{zT}$ are the counts of calibration survey galaxies in both redshift bin $z$ and superphenotype $T$.  These counts must be drawn from a calibration field with high-quality redshift assignment.  A shortcoming of this step is that the Dirichlet distribution assumes no correlation between fluctuations in distinct $z$ bins (except those induced by the $\sum f=1$ constraint), whereas we know that sample variance does have a finite correlation length in $z.$  Figure~\ref{Fig:SVCorrelation} shows inter-bin density correlations are $<0.1$ at $\Delta z\ge0.05$ for $z\lesssim 1$ in the COSMOS2015 fields.  Our method will hence not sample the large-scale structure faithfully on scales below $\Delta z \approx 0.05.$

Finally, we wish to draw a sample of the $f_t^{zT}$ probabilities, \ie\ to distribute the probability of a given redshift bin among the phenotypes $t$ at that redshift.  Because the superphenotypes $T$ are nearly disjoint in redshift, we opt to simplify this process at some small loss in accuracy by summing over superphenotypes, and using $f_t^z \equiv p(t |z)$  instead of $p(t | z,T)$ by executing the following step for each redshift bin $z$:
\begin{enumerate}[leftmargin=1cm]
  \setcounter{enumi}{\value{enumCount}}
  \item $f_{tz} \leftarrow \mathrm{Dir}(M_{tz}) \: f_z$; with $f_z = \sum_T f_{zT}$
  \setcounter{enumCount}{\value{enumi}}
\end{enumerate}
In this case $M_{tz}$ are the galaxy counts in joint redshift-phenotype space, which again requires a high-quality redshift calibration field.

Note that there is no sample-variance inflation factor $\lambda$ in the final sampling step, since we are assuming that the sample variance is strictly a redshift phenomenon---within a redshift bin, the phenotype assignments are assumed to have only shot noise.  This also means that our sampling method is thus assuming that all phenotypes have the same bias.

In summary, this sampling method for $f_{zt}$ works by splitting the redshift and phenotype parts of the problem and takes advantage of our knowledge of sample variance across redshift from the previous Sections by inflating the variance in the redshift axis.
This ``3sDir'' method reduces to the basic Dir method described above when  $\lambda_T=1$ at all $T.$

\begin{figure*}
\centering
	\includegraphics[width=1.05\textwidth]{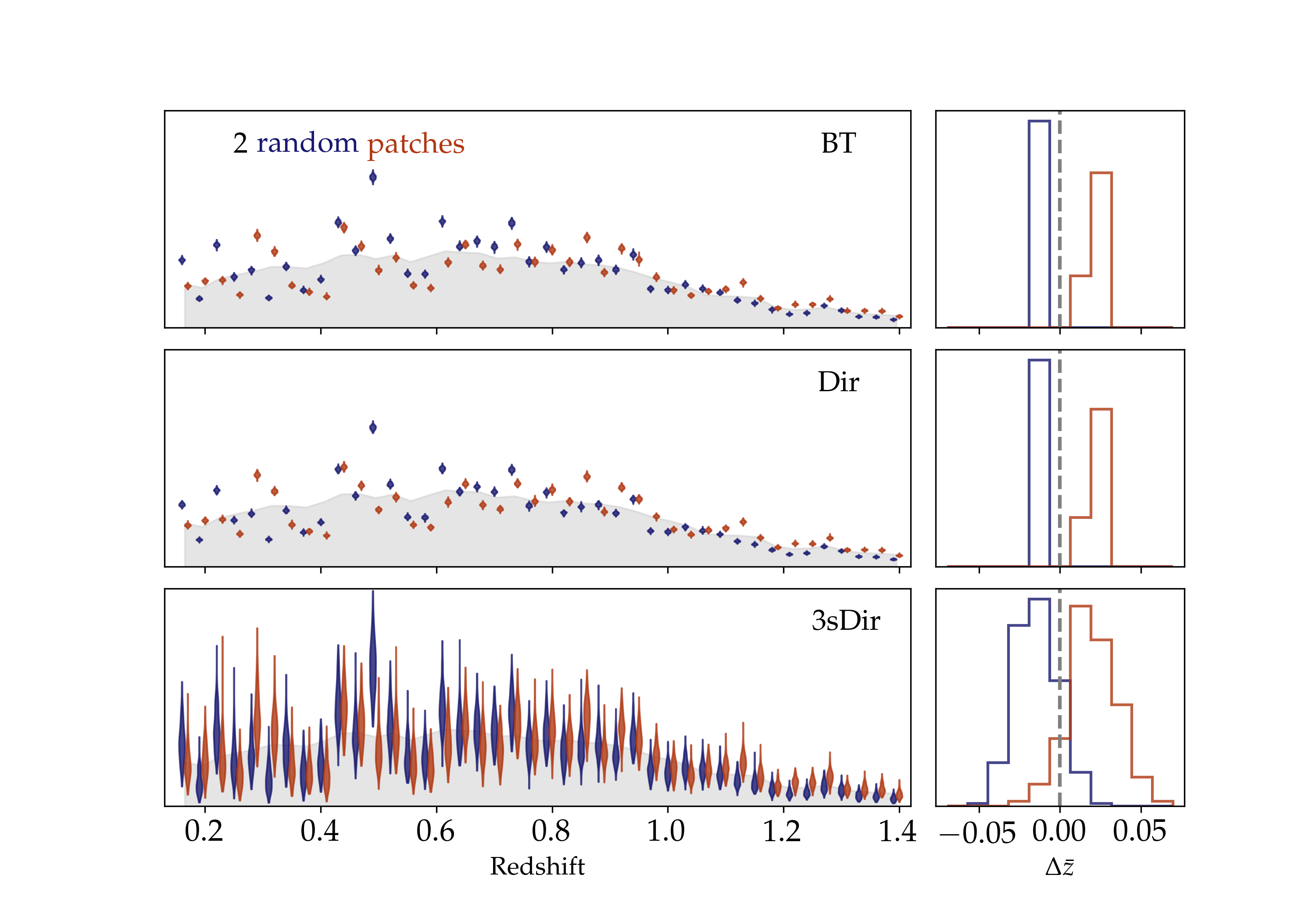}
	\caption{Two random patches chosen from the simulation. We show, for each sampling method (BT, Dir and 3sDir), the sampling of the redshift distribution of the patches (left) and its mean redshift relative to the true mean (right). The true redshift distribution of the population is shown in grey, and the true mean redshift as the vertical dashed line.  Clearly only the 3sDir method generates samples with uncertainties large enough to include the truth values. }
\label{fig:nzs}
\end{figure*}

The splitting into $z$ and $t$ samplers enables lower noise for the case where one has more calibration data with high quality photometry than with high-quality redshift information.  One could generalize this method to allow, for each phenotype, combining calibration fields with differing levels of redshift uncertainty (\eg\ spectroscopic vs many-band photometric redshifts). This is left for future work.

\section{Validation in simulations}
\label{sec:results}

\begin{figure*}
\centering
	\includegraphics[width=0.9\textwidth]{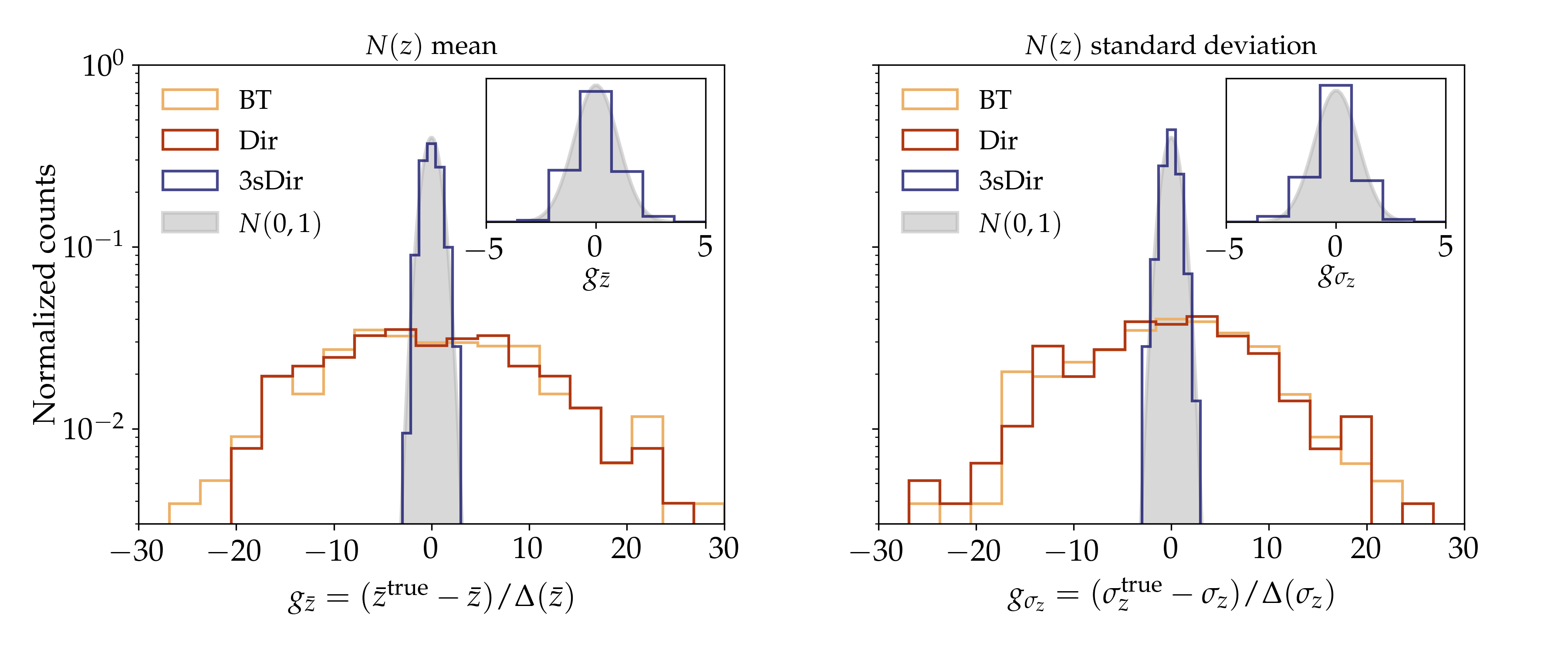}
	\caption{\textit{(Left panel):} Pull distribution for the mean of the redshift distribution $\bar{z}$ for each of the three sampling methods. If uncertainties are well behaved, pulls should approach a Gaussian $N(0,1)$ (in grey). Inset shows a zoom in for the 3sDir method. \textit{(Right panel):} Same as in left panel, but for the standard deviation of the redshift distribution. }
\label{fig:pull}
\end{figure*}

We apply the three sampling strategies described in Section \ref{sec:method} (BT, Dir and 3sDir) to the sky patches defined on the MICE simulation and we test their accuracy in reproducing the true uncertainties from sample variance. We use the 247 patches of the simulation (described in \S\ref{sec:sims}). For any single patch, each of the three methods is able to produce a number of realizations of the $f_z=p(z)$ function, and we can test if those realizations accurately describe the uncertainties of the redshift distribution coming from that patch. For the 3sDir method, we utilize the sample variance theory characterization of Section \ref{sec:theory} as an input to the method.    

As a first test of the sampling methods, we randomly choose two patches from the simulation, and produce realizations of $f_z$ from them using the three different sampling schemes (BT, Dir and 3sDir). In Figure~\ref{fig:nzs}, the three panels on the left column show the true redshift distribution of the galaxy population (in grey) and violin plots for the sampling of the redshift distribution from the two random patches by the three different methods, BT, Dir and 3sDir, in the upper, center and lower panels, respectively. For the first two cases, bootstrap and basic Dirichlet, the uncertainties coming from the sampling are underestimated and do not cover the true redshift distribution of the population, and both methods perform similarly. The 3sDir method does a much better job, with the sampling containing the true distribution. The panels in the right column show the distribution for the mean of each of the samples of the redshift distribution produced by the three schemes, minus the true mean redshift of the population. The BT and Dir methods yield very tight distributions for the mean redshift difference, but not containing the truth value of zero. In contrast, the 3sDir method shows wider distributions of the mean redshift, containing the truth for both patches.

A more quantitative test of the fidelity of the sampler of $f_z$ is to compute the \textit{pull} distribution of the mean-$z$ errors. For each simulated patch $i$, we generate 100 samples $j=1\ldots100$ of $f_z,$ then calculate the mean redshift  $\bar{z}^{ij}$ for each distribution.  For a given patch we can then calculate a mean error $\bar{z}^i$ over the samples and a standard deviation $\Delta(\bar{z})^i$.  The ``pull'' $(z_{\mathrm{true}} - \bar{z}^i)/\Delta(\bar{z})^i$ for the 247 patches should approach a Normal distribution $\mathcal{N}(0,1)$ if the sampler is properly estimating the total uncertainty in the redshift distribution, and is unbiased, and if the uncertainties are Gaussian.
Similarly, we can compute the standard deviation $\sigma_z$ of the redshift distribution for each sample from each patch; and we can plot the pull distribution of this summary statistic of $p(z).$

Figure \ref{fig:pull} shows the pull distribution of both $\bar{z}$ and $\sigma_z$ for each of the three sampling methods.  As expected from the test in Figure \ref{fig:nzs}, both the bootstrap and the basic Dirichlet method perform poorly in the comparison with a Gaussian $\mathcal{N}(0,1)$. On the other hand, the 3sDir method performs very well in that comparison (see inset panels in Figure \ref{fig:pull}), demonstrating that the method properly captures the sample variance uncertainties in the mean and width of the redshift distributions. Both the mean and the width of redshift distribution are key properties for galaxy clustering and cosmic shear studies.

\begin{figure}
\centering
	\includegraphics[width=0.5\textwidth]{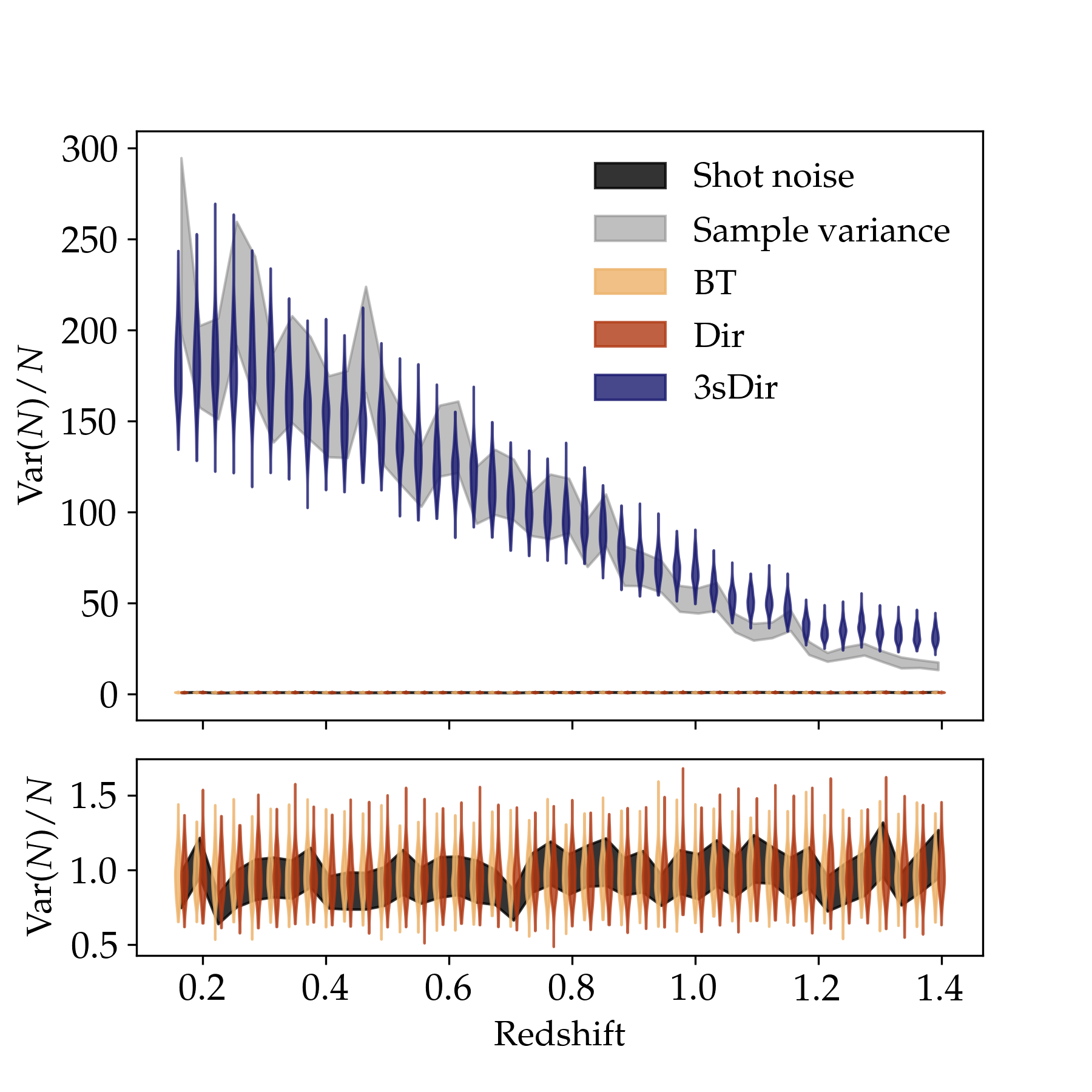}
	\caption{Normalized variance in the samples of the redshift distribution as a function of redshift, $\mathrm{Var}(N_z)/N_z$, for the three sampling methods (violins). Sample variance and shot noise estimates are shown in grey and black.}
\label{fig:varz}
\end{figure}

Another quantity that we would like our sampling scheme to reproduce well is the (normalized) redshift variance $\mathrm{Var}(N_z)/N_z$ as a function of $z$. That is, we would like the samples to span the same range of $N_z$ as is expected from the combination of shot and sample variance.
As seen in Figures \ref{fig:uncertainties} and \ref{Fig:SVFullTheory}, this quantity has significant redshift dependence in both simulations and theory.  Figure \ref{fig:varz} shows the 3sDir method succeeds in producing samples that follow the expected redshift dependence in normalized variance. On the contrary, the BT and Dir methods yield variances consistent with pure shot noise, far below the sample variance level.

With this we have shown how the proposed 3-step Dirichlet sampling scheme can reproduce the uncertainties from sample variance in the redshift distribution, its mean, its width, and its variance as a function of redshift.

\section{Application to the COSMOS2015 data set}
\label{sec:cosmos}

We now apply the methods of the previous sections to real data, namely the COSMOS2015 catalog \citep{Laigle2016}, which provides high quality redshift estimates at high completeness over the color-magnitude space down to faint magnitudes.  COSMOS2015 has played a key role in the redshift characterization of many past and current cosmological analyses using lensing surveys \citep{Bonnett2015,Hoyle2017,Hildebrandt2018,Hikage2018,Hamana2019}.  We produce resamplings of this catalog (reweightings) that realize uncertainties from both shot noise and sample variance.  

\subsection{Redshift estimates}
\label{sec:cosmos_redshifts}

The COSMOS2015 catalog from \citet{Laigle2016} provides photometry in 30 different UV/visible/IR bands, and probability distribution functions (PDFs) $p(z)$ for the redshift of each galaxy based on this photometry using the LePhare template-fitting code \citep{Arnouts1999,Ilbert2006}. Due to the extensive photometric coverage of the catalog, the redshift performance is very good: For bright galaxies, the typical $p(z)$ widths are $\sim 0.01(1 + z)$, with outlier rates around 0.5\%, while for fainter, high-redshift galaxies it goes to $p(z)$ widths of $\sim 0.023(1 + z)$ and outlier rates of about 13\%. For the results in this work, we use the redshift estimates of \texttt{ZMINCHI2}, which is the 30-band photometric redshift point prediction corresponding to the the minimum $\chi^2$ fit between fluxes and templates, and throughout this Section, we use a redshift binning of width 0.05 between redshift 0.06 and 5.01, what makes a total of 99 redshift bins.  

\subsection{Phenotype characterization}
\label{sec:t_cosmos}

Similar to our treatment of the MICE2 simulations, phenotypes for COSMOS2015 are defined as cells in a Self-Organizing Map (SOM) which is trained on photometric data. In this case, we use the following photometric bands: $\{u,B,V,r,ip,zp,,J,H,Ks\}$. We limit our sample to galaxies with $\mathrm{mag}(ip)<25.5.$ Quality cuts on the data\footnote{We select objects with \texttt{TYPE}==0, \texttt{FLAG\_PETER}==0, \texttt{B\_IMAFLAGS\_ISO}==0, \texttt{B\_FLUXERR\_APER3}>0, where \texttt{B} runs over all photometric bands we use.} yield a total of 305,835 galaxies placed in the SOM. For classifying the COSMOS2015 catalog, we make substantial changes to the SOM algorithm with the purpose of improving the classification of galaxies of modest $S/N$, and to allow magnitude information (not just colors) to be used in redshift estimation.  These improvements to the SOM algorithm are detailed in Appendix A.    For present purposes it suffices to note that the COSMOS2015 galaxies are each assigned to one of $64\times64=4096$ phenotypes defined by a SOM cell.

\begin{figure}
\centering
	\includegraphics[width=0.5\textwidth]{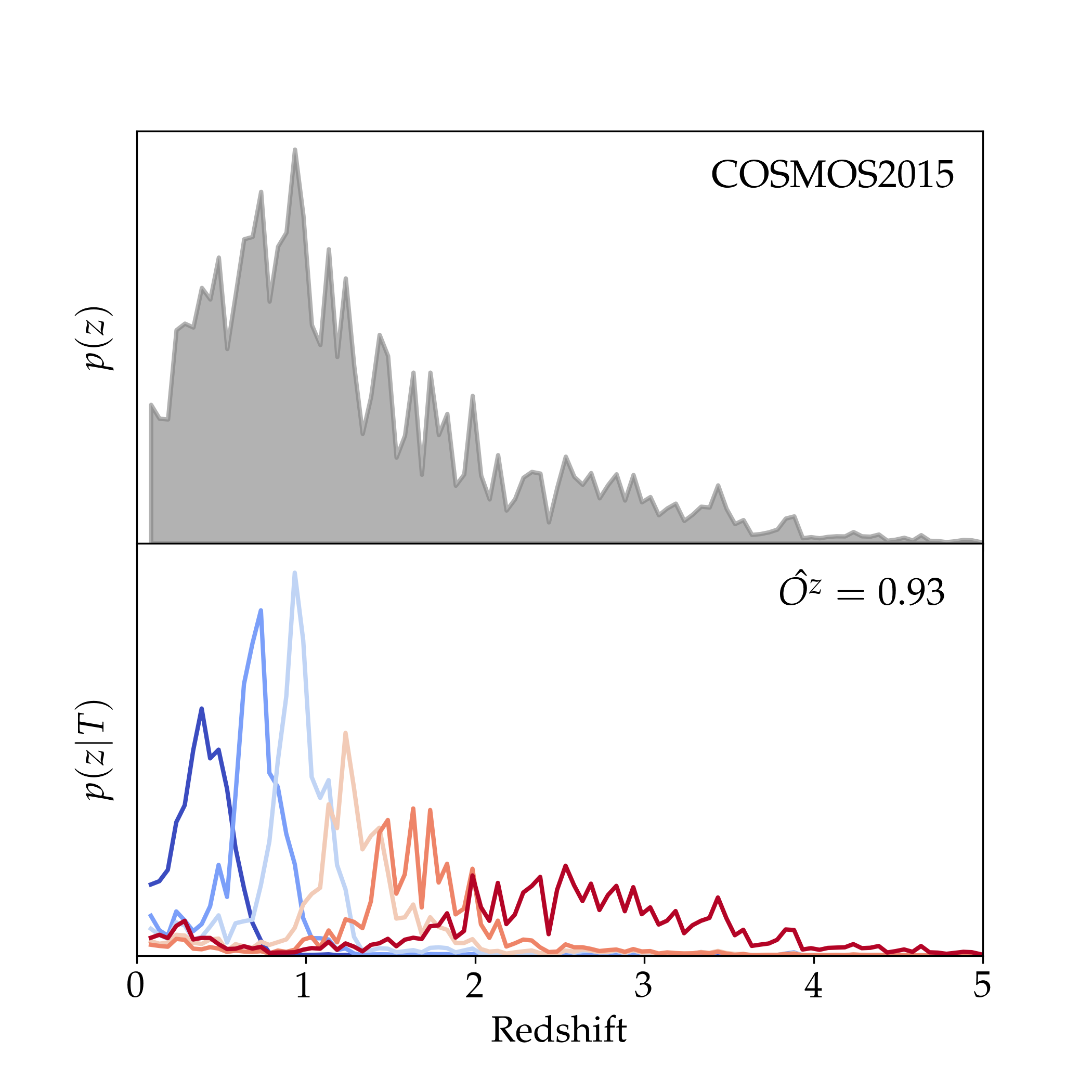}
	\caption{\textit{(Upper panel)}: Redshift distribution for the COSMOS2015 galaxies used in this Section. \textit{(Lower panel)}:  Redshift distributions for the different superphenotypes made for the COSMOS2015 data. This is similar to Figure \ref{fig:nzTs} in the simulation. The overlap metric of Equation (\ref{eq:overlap}) for this case is $\hat{O^z} = 0.93$.}
\label{fig:nzTs_cosmos}
\end{figure}

As described for the MICE simulation in \S\ref{sec:phenotypes} and Figure \ref{fig:nzTs},
we divide the COSMOS2015 phenotypes into 6 \textit{superphenotypes} by partitioning a list of
all phenotypes ordered by their mean redshifts.
Due to the large redshift range of the COSMOS2015 galaxy sample, we create 4 equally-spaced superphenotypes between redshift 0 and 1.75, and then two-equally spaced high-redshift superphenotypes between redshift 1.75 and 5. The phenotypes resulting from this prescription are shown in Figure \ref{fig:nzTs_cosmos}, and they have an overlap metric of 0.93, computed as in Equation (\ref{eq:overlap}).    

\subsection{Sample variance estimation} 
\label{sec:sv_cosmos}
We produce a theoretical estimate for sample variance (SV) using the methods of Sec.~\ref{sec:theory}.
For simplicity we fix all cosmological parameters to their best fit Planck 2018 values, as obtained by fitting the $\Lambda$CDM model to the combination of the Planck measurements of CMB temperature and polarization, Planck CMB lensing reconstruction and Baryon Acoustic Oscillations measurements~\citep{Aghanim:2018eyx}.
The crucial part in the prediction of SV is the modeling of galaxy bias. As in Sec.~\ref{SubSec:SimulationComparison} we first extract bias from the COSMOS data by fitting the amplitude of the measured correlation function at different redshifts.
Then we use the halo model and we fit its parameters to reproduce the observed COSMOS redshift distribution so that we can predict the value and redshift evolution of bias.
We find that these two procedures produce results that match to satisfactory accuracy.

After the recipe for galaxy bias is obtained we proceed with the estimate of SV following Eq.~\ref{Eq:TheorySV} with a smoothing filter obtained assuming circular survey geometry matching the area of the COSMOS field.
In this case, since we are not matching a simulation, we produce an estimate of the full sky SV, taking into account contributions coming from all scales bigger than the COSMOS patch.
The resulting SV qualitatively follows the previous results in Fig.~\ref{Fig:SVFullTheory}, decreasing as redshift increases.

\begin{figure}
\centering
\includegraphics[width=\columnwidth]{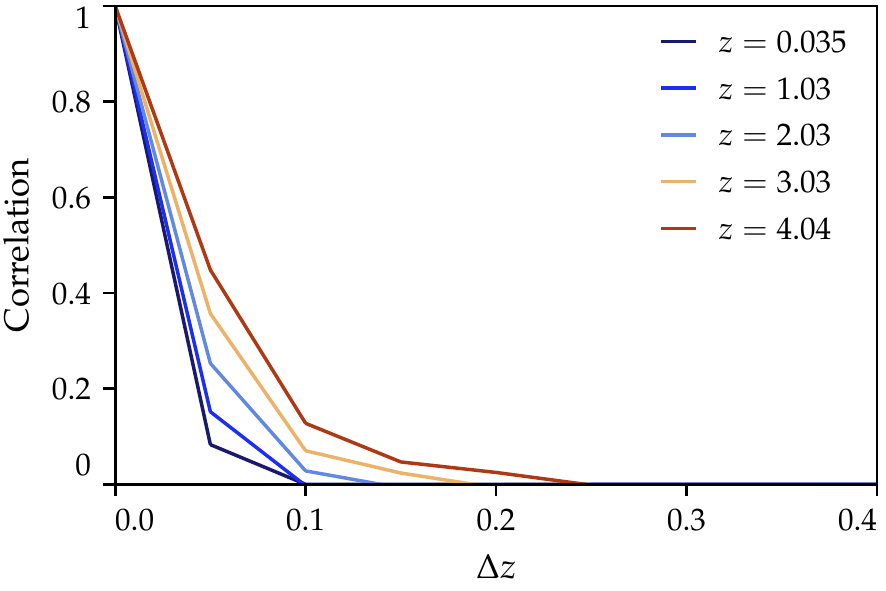}
\caption{ \label{Fig:SVCorrelation}
Theory prediction of sample correlation as a function of redshift separation for the COSMOS sample.
Different colors represent different redshifts, as indicated in the legend.
}
\end{figure}

The cross-correlations between different redshift bins, ${\rm Corr} (\hat{\Delta}_z,\hat{\Delta}_{z'}) = {\rm SV}(z,z')/[\sqrt{{\rm SV}(z)}\sqrt{{\rm SV}(z')}]$,
are independent of bias modeling and depend on the long modes along the line of sight. 
Figure~\ref{Fig:SVCorrelation} plots predicted correlation coefficients as a function of redshift separation and at different redshifts.
At low redshift ($z\sim 1$) sample correlation decays quickly in redshift and becomes negligible at a redshift separation of $\Delta z\sim 0.1$. On the other hand the decay is slower at higher redshift.
The correlation predictions can be used to choose appropriate smoothing functions for the resampled COSMOS2015 redshift distributions, a topic that we will defer to future work.

\begin{figure*}
\centering
	\includegraphics[width=1.\textwidth]{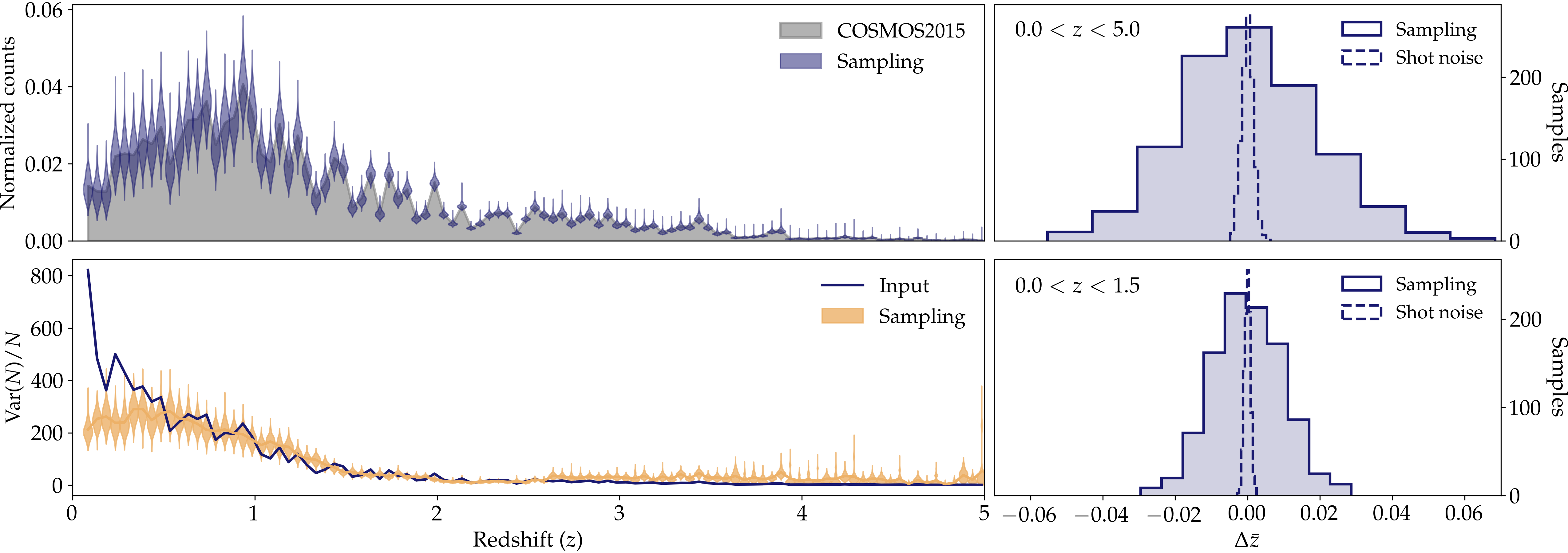}
	\caption{\textit{(Upper-left panel)}: Redshift distribution for the COSMOS2015 galaxies, and one-sigma error bars for the samples produced in this work, which include uncertainties from shot noise and sample variance. \textit{(Lower-left panel)}: Input normalized redshift variance to the sampling method, together with resulting normalized redshift variance for the samples produced. Uncertainties on the latter come from bootstrap resampling. \textit{(Right panels)}: Distributions of mean redshift for the redshift distribution samples produced, both for the whole redshift range (upper), and for a subset at low redshift ($z<1.5$) (lower). We also show the distributions corresponding to shot noise only, without sample variance, for comparison.}
\label{fig:sampling_cosmos}
\end{figure*}

\subsection{Sampling} 
We can now apply the 3sDir sampling method from Section \ref{sec:method} and produce realizations of the COSMOS2015 redshift distribution that include uncertainties from sample variance and shot noise. For that, we will use as inputs both the phenotype and superphenotype definitions of \S\ref{sec:t_cosmos} and the theory sample variance estimation of \S\ref{sec:sv_cosmos}. We will apply the sampling method as described in \S\ref{sec:3sDir}, and we will assign the minimum sample variance contribution to the two high-redshift superphenotypes defined in \S\ref{sec:t_cosmos}. Figure \ref{fig:sampling_cosmos} shows the result of the 3sDir sampling method applied to the COSMOS2015 data sample. The upper left panel shows the COSMOS2015 redshift distribution together with the mean and standard deviation of the samples in each redshift bin. The lower left panel shows the normalized redshift variance that was input to the sampling method, compared to the normalized redshift variance of the the samples, showing generally good agreement. The right panels show the distributions of the mean redshift difference between the samples and the input COSMOS2015 data, for the entire redshift range and for a subset at lower redshift. In those panels, we also show the distribution corresponding to shot noise only, without sample variance, for comparison. For the entire redshift range ($0<z<5$) the 3sDir method shows a mean redshift dispersion of 0.019 (0.002 for shot noise only), while for the lower redshift subsample ($0<z<1.5$) it goes down to 0.0097 (0.0008 for shot noise only).

We release with this paper a set of 1000 realizations of the COSMOS2015 $n(z)$ produced by the 3sDir method, which anyone can use to propagate the shot and sample variance of COSMOS2015 into their own analyses.  This is presented as a table with 305,835 rows (one for each COSMOS galaxy passing the quality and $ip<25.5$ cuts); there are columns for the COSMOS2015 ID number of the galaxy, its \texttt{ZMINCHI2} redshift, and the bin number $z$ and phenotype $t$  to which it is assigned in the SOM.  There are then 1000 columns labelled by index $j$ containing
weights to apply to the galaxies to realize the $j$th 3sDir sample.  The weights are equal to $w_{ij} = f_{tz}^j*M_{\rm tot}/M_{tz},$ where $f_{tz}^j$ is the value of $f_{tz}$ 3sDir sample for the $tz$ bin to which galaxy $i$ belongs; $M_{\rm tot}=305,385,$ and $M_{tz}$ is the number of COSMOS2015 galaxies falling into the $tz$ bin. The file containing the table described above can be downloaded in \texttt{FITS} format from this link\footnote{https://cosmos2015resampling.shortcm.li/sbrdMS}.

To use the 3sDir resamplings, a user makes any particular cuts they desire to the galaxy sample; defines their own redshift bins; and then makes a histogram of the galaxies in these bins, weighting by the $w_{ij}.$  Each column will yield an independent sample of the $n(z).$

\section{Summary and conclusions}
\label{sec:con}

Redshift uncertainties have become one of the leading contributions to the systematics budget of imaging galaxy surveys. In particular, they are important enough to bias cosmological constraints in a way that different surveys can be at tension between themselves and with results from other probes such as the CMB, and hence it is crucial to properly include them in cosmological analyses. 

Such uncertainties in the redshift distribution of a galaxy population arise mostly because of the limited knowledge of the color-redshift relation, which usually comes from prior knowledge in small patches of the sky known as calibration samples. In those samples, galaxy redshifts are known either through spectroscopy or from high-quality photometry, and they have associated uncertainties regarding those estimations. In addition, due to the small size of calibration fields, sample variance from large-scale structure becomes an important source of uncertainty \citep{Cunha2012}. In this paper, we have studied in detail the impact of sample variance and shot noise in the uncertainties associated to the redshift distributions of calibration samples, both from theory and $N$-body simulations, and we have proposed a new scheme to produce realizations of redshift distributions including those uncertainties.        

In addition, using the techniques described above and a dedicated self-organizing map (SOM) algorithm, we have applied the scheme to the COSMOS2015 data sample, producing a theory estimate of the sample variance contribution for that particular data set, and then generating (and making public) 1000 realizations of its redshift distribution that include the effects of shot noise and the estimated sample variance. From those realizations, we compute an uncertainty in the mean redshift of the COSMOS2015 population to be around 2\% ($\simeq$1\% for $z<1.5$). That uncertainty is comparable to the redshift uncertainties estimated in cosmological analyses that have used COSMOS2015 as redshift prior (\eg~\citealt{Hoyle2017}), which highlights the importance of correctly propagating it into cosmological constraints from galaxy surveys.  

In summary, this work introduces three main advances to the problem of propagating redshift uncertainties into the analysis of imaging galaxy surveys: 
\begin{enumerate}
\item A theoretical approach to estimating the sample variance contribution to the uncertainty in the redshift distributions of calibration fields. A theoretical estimate has several advantages over a direct estimation from $N$-body simulations, such as the unlimited redshift range and the possibility of handling different modeling effects such as cosmology dependence.     
\item A sampling scheme, based on the Dirichlet distribution, to produce realizations of the redshift distribution of a calibration field given estimates of shot noise (which can be taken directly from the data) and sample variance (given by (i) or by simulations). The realizations are able to correctly reproduce the sample variance impact in the uncertainty in the mean and standard deviation of redshift distributions, and the redshift trend of the number counts uncertainty.    
\item A new SOM algorithm designed to improve the phenotype classification of galaxies of modest $S/N$, and to allow magnitude information (not just colors) to be used in redshift estimation.
\end{enumerate}

These new elements of redshift calibration presented in this work provide an appropriate way of including sample variance uncertainties into the redshift uncertainty budget, which was an important missing piece in past analyses. This work will therefore contribute to an improved $N(z)$ characterization in current and future real survey data, which is a key part of the overall systematic uncertainties in future weak lensing and galaxy clustering analyses.

\section*{Acknowledgements}

The authors thank Cyrille Doux, Lucas Secco, Daniel Gruen, Justin Myles, Joe DeRose and Alexandra Amon for helpful conversations about this topic.  
This work was supported by grants AST-1615555 from the US National
Science Foundation, and DE-SC0007901 from the US Department of Energy. 
MR is supported in part by NASA ATP Grant No. NNH17ZDA001N, and by funds provided by the Center for Particle Cosmology. 
Work at Argonne National Lab is supported by UChicago Argonne LLC,Operator of Argonne National Laboratory (Argonne). Argonne, a U.S. Department of Energy Office of Science Laboratory, is operated under contract no. DE-AC02-06CH11357.

 %%%%%%%%%%%%%%%%%%%%%%%%%%%%%%%%%%%%%%%%%%%%%%%%%%


\bibliographystyle{mnras}
\bibliography{./library}

 %%%%%%%%%%%%%%%%%%%%%%%%%%%%%%%%%%%%%%%%%%%%%%%%%%

 %%%%%%%%%%%%%%%%% APPENDICES %%%%%%%%%%%%%%%%%%%%%

 \appendix

 % Appendix on equivalence to Newman etc.
\section{COSMOS2015 SOM}
\label{sec:som}

In this Appendix we describe the self-organizing map (SOM) created with
COSMOS2015 data, which is used in Section \ref{sec:cosmos} to define
the phenotypes of that sample. The use of SOMs to discretize the
galaxy color/magnitude space has been discussed extensively in
previous works \citep[\eg][]{Masters2015, Speagle2019, Buchs2019}, and we refer the
reader to those works and references therein for more details about
the standard SOM algorithm.  Here we will describe the ways in which
the algorithm we apply to COSMOS2015 differs from previous
implementations.

  \begin{figure}
\centering
	\includegraphics[width=0.5\textwidth]{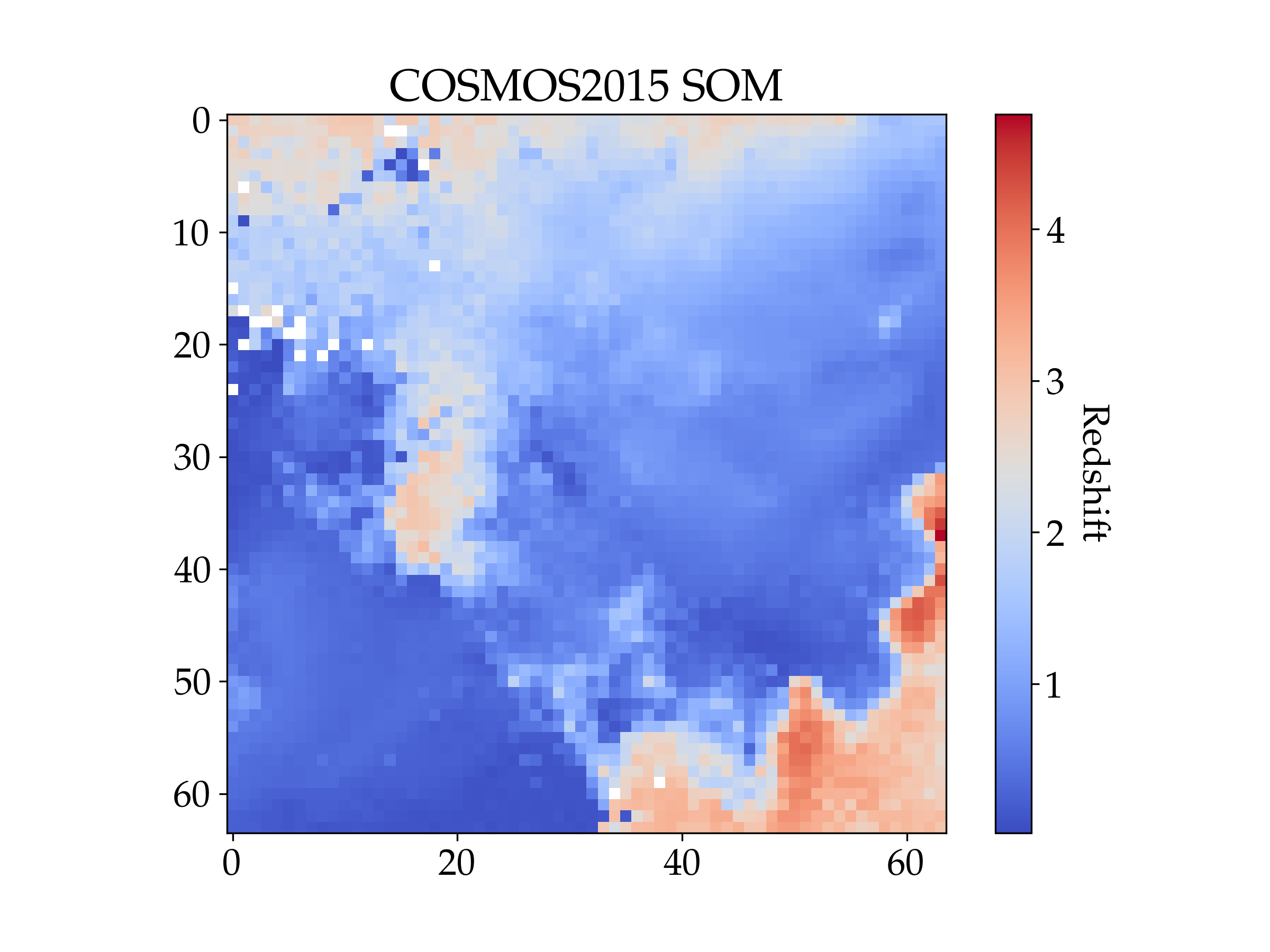}
	\caption{The COSMOS2015 SOM, coloured by the mean redshift in
          each cell. White cells have no members.}
\label{fig:som_cosmos}
\end{figure}

The gist of the algorithm is as follows: first, we have a
population of objects (galaxies) labelled by $i=1,2,\ldots,N_{\rm obj},$ each
of which has a measured feature set ${\bf F}_i$ in a $d_F$-dimensional
space consisting of, in our
cases, the fluxes of the object in the $u,B,V,r,ip,zp,,J,H,$ and $Ks$
bands.  In our case we also have measurement uncertainties $\bm{\sigma}_i$
corresponding to each ${\bf F}_i$---the standard SOM algorithm does
not allow for measurement errors in the training set, so this is one
aspect we wish to address.

The SOM is defined by discrete set of points ${\bf
  x}_c$ in an $d_{\rm SOM}$-dimensional space. Typically these are
placed in a regular grid in $d_{\rm SOM}=2$ dimensions for photo-$z$
applications.  Each such ``cell'' is assigned a location in the 
feature space, ${\bf F_c}={\bf F}({\bf x}_c),$ known as the
``weights'' for the cell in the SOM literature.  The final step in
specifying the SOM is to define some distance measure $r({\bf
  F}_c,{\bf F}_i,\bm{\sigma}_i)$ between a cell and an object.
Typically this is taken as the Euclidean metric $|{\bf F}_c-{\bf
  F}_i|,$ but this is not required.  There is in fact no need for $r$
to satisfy the conditions of a metric, nor even to be symmetric in its
arguments.  Below we will describe a distance function which is much
more desirable in the photo-$z$ application than the Euclidean 
distance.

An arbitrary object $i$ is assigned to a SOM cell by simply selecting
the cell $c$ which minimizes $r({\bf F}_c,{\bf
  F}_i,\bm{\sigma}_i).$  The SOM literature calls $c$ the ``best
matching unit'' (BMU) for the object.

The algorithm for assigning the weights $F_c$ to the cells using
a sequence of training objects $F_j,$ $j=1,2,\ldots,N_{\rm train}$
from the population is:
\begin{enumerate}
  \item Initialize the weights to a set ${\bf F}_c^{(0)}$ in the
    regime spanned by the data.
   \item For each training point $j$, find its BMU, then alter
     the full set of weights according to
     \begin{equation}
       {\bf F}^{(j)}_c =  {\bf F}^{(j-1)}_c + A(j) H({\bf x}_c- {\bf
         x}_{\rm BMU}, j) \, \bm{\Delta}({\bf F}_c^{(j-1)},{\bf
           F}_i,\bm{\sigma}_j.)
     \end{equation}
   \end{enumerate}
In the training phase, we have a shift in the feature space,
${\mathbf\Delta}$, to draw the BMU weights toward the training
object. In the standard algorithm $\bm{\Delta}$ is simply ${\bf
  F}_i-{\bf F}_c^{(j-1)},$ but we will alter this.
The function $H$
describes a neighborhood around the BMU in the SOM space which is
dragged toward the training point, which contracts with advancing
iterations $j.$  Finally there is an overall learning rate $A$ which
also typically decreases as training proceeds.

We adopt the functional forms for $A$ and $H$ given by
\citet{Speagle2019}.  We adopt a regular grid of ${\bf x}$ in 2 dimensions
for our cells, as have other photo-$z$ applications, in our case a
$64\times64$ array.  We do \textit{not} adopt period boundary conditions
in SOM space, because there is no natural periodicity in the feature
space of galaxy colors and magnitudes.

The main alterations we make to previous methods are to the distance
function $r$ and the training shift $\bm{\Delta}.$  Previous works have
struggled with the choice between using fluxes, magnitudes, and/or
colors as the elements of the feature space.  Fluxes are the natural
space in the sense that measurement errors are nearly Gaussian in flux
space, making a Euclidean distance the best match to a (log)
probability in flux space.  It is also true, however, that galaxy
\textit{colors}---\ie\ ratios of fluxes, or differences in log(flux)---are more sensitive indicators of
redshift than are fluxes, making a Euclidean metric in color a better
way to group galaxies into cells of common redshift.  For this reason,
most SOM-based photo-$z$ techniques use colors as features.
But this method
fails catastrophically when one or more bands have modest or low
$S/N$ (or when observed fluxes are negative!), as color becomes wildly
uncertain, and the measurement noise can come to dominate the choice
of BMU.  We want our functions $r$ and $\bm{\Delta}$ to respect the meaning
of the error bars, and not use unreliable information to classify
objects or train the SOM. A further shortcoming of pure color-based SOMs is that
redshift is known to very with flux (or magnitude) at fixed color, as
demonstrated by \citet{Speagle2019} and \citet{Masters2019}.  We therefore
want a distance function with sensitivity to overall flux or
magnitude level, but much more weakly than to color.

The solution we find is as follows.  First, we use fluxes as our
features, so that negative measured fluxes can be treated properly.
But we force the cells to have positive weights (fluxes), since we
consider the cell weights to be noiseless galaxy ``templates.''
We then define a nominal distance function to be a sum over the
elements of the flux vector (indexed by band $b$) as follows.  First
we convert the object and cell fluxes into units of signal-to-noise
ratio (SNR), specifying a maximum for the object SNR as a means of
softening the specificity of the cells:
\begin{align}
s_{ib} & \equiv {\rm max}\left(\sigma_{ib},F_{ib}/{\rm SNR}_{\rm
          max}\right) \\
 \nu_{ib} & \equiv F_{ib}/s_{ib} \\
 \nu_{cb} & \equiv F_{cb}/s_{ib}. 
\end{align}
Next we define a weighting function that will be used to transition
from the high- to low-SNR regimes:
\begin{equation}
  w_{ib} = e^{2(\nu_{ib}-4)}.
\end{equation}
Now we define a distance function
\begin{equation}
  \tilde r({\bf F}_c,{\bf F}_i,\bm{\sigma}_i) =
  \sum_b \left[ \frac{\asinh\nu_{cb} + w_{ib}\log{2\nu_{cb}}}{1+w_{ib}}
      - \asinh \nu_{ib}\right]^2 \left(1 + \nu^2_{ib}\right)
  \end{equation}
which has the desirable properties of approaching Euclidean in
log-flux at high SNR, and Euclidean in linear flux at low SNR,
weighting each band by its SNR (up to a maximum),
and
monotonically increasing away from ${\bf F}_i={\bf F}_c.$

The final step is to make the cells ``fuzzy'' in overall flux level,
in the sense of \citet{speagle2015}, by allowing the cell fluxes to be
scaled by an overall constant $e^s$:
\begin{equation}
r({\bf F}_c,{\bf F}_i,\bm {\sigma}_i) = \inf_s \left[\tilde
  r(e^s{\bf F}_c,{\bf F}_i,\bm{\sigma}_i)
   + \frac{s^2}{\sigma_s^2} \right]
\end{equation}
The cell's fluxes are thus rescaled to find the minimal distance to
the object's fluxes, subject to a penalty that is quadratic in the log
of the scaling factor.  The parameter $\sigma_s$ determines,
essentially, the width in magnitudes of the smeared cells.  We thus
realize a distance that can be quite sharp in any color while being
broad in overall flux.

We will not detail the shift function ${\mathbf\Delta}$ here, just
noting that it operates as a simple
shift in log-flux space when SNR is high in a given band, but produces
no shift during training in bands where the difference between the
flux of the training object and the flux of the cell have less than
$2\sigma$ significance.

The COSMOS2015 SOM is created using ${\rm SNR}_{\rm max}=50,$ which
essentially sets the sensitivity to color at $\approx0.02$~mag; and
with $\sigma_s=0.4,$ which sets the sensitive to overall flux at
$\approx0.4$~mag. 
Figure~\ref{fig:som_cosmos} shows the mean redshift of COSMOS2015
galaxies assigned to each SOM cell, plotted across the SOM space ${\bf
  x}.$  The redshift standard deviation of galaxies assigned to each cell has a median of 0.21. Keep in mind that the SOM was constructed without regard to
redshift. The smooth behavior of redshift across the SOM shows that
the variation of redshift in the 9-band flux space is reasonably well
traced by the embedded 2d manifold defined by the SOM.

 %%%%%%%%%%%%%%%%%%%%%%%%%%%%%%%%%%%%%%%%%%%%%%%%%%

 % Don't change these lines
\bsp	% typesetting comment
\label{lastpage}
\end{document}